\documentclass[preprint,5p,times,authoryear]{elsarticle}
\journal{Physics of the Dark Universe}

\usepackage[utf8]{inputenc}	
\usepackage[T1]{fontenc} 		
\usepackage{lmodern}				
\usepackage{microtype}			

\usepackage{graphicx}				
\usepackage{amsmath}				
\usepackage{amssymb}				
\usepackage{siunitx}				
\usepackage{csquotes}				
\usepackage{hyperref}				
\usepackage{cleveref}				
\usepackage{acro}						
\usepackage{caption}				
\usepackage{url}						
\usepackage[all]{nowidow}		
\usepackage[draft]{changes}	

\usepackage{xpatch}
\usepackage{ulem}

\usepackage{aas_macros}
\usepackage{elsfix}
\hypersetup{
	colorlinks,
	linkcolor={red!50!black},
	citecolor={magenta},
	urlcolor={blue!80!black}
}

\renewcommand{\vec}[1]{\mbox{\boldmath$#1$}}

\DeclareCaptionFormat{cont}{#1 (cont.)#2#3\par}

\DeclareSIUnit \c 			{\ensuremath{\mathit{c}}}
\DeclareSIUnit \parsec 	{pc}
\DeclareSIUnit \eV 			{eV}
\DeclareSIUnit \eVcc		{eV\per\c\squared}
\DeclareSIUnit \Msun 		{M_\odot}

\setauthormarkup{}
\definechangesauthor[name={Andreas}, color=purple]{ak}
\definechangesauthor[name={Carlos}, color=purple]{ca}
\definechangesauthor[name={Jorge}, color=purple]{jr}

\crefname{appendix}{}{}
\Crefname{appendix}{}{}
\newcommand{\diff}{\mathop{}\!\mathrm{d}}							
\newcommand{\tdiff}[2]{\diff{#1}/\diff{#2}}						

\newcommand{\abs}[1]{\left|#1\right|}									

\DeclareAcronym{rho-crit}{
  short = \ensuremath{\rho_{\rm cr}},
  long  = the critical density of the Universe,
  class = symbol
}

\DeclareAcronym{lambda-B}{
  short = \ensuremath{\lambda_{\rm B}},
  long  = de Broglie wavelength,
  class = symbol
}

\DeclareAcronym{r-half}{
  short = \ensuremath{r_{\rm half}},
  long  = half light radius,
  class = symbol
}

\DeclareAcronym{r-last}{
  short = \ensuremath{r_{\rm last}},
  long  = last observed data point,
  class = symbol
}

\DeclareAcronym{r-300}{
  short = \ensuremath{r_{300}},
  long  = radius at $\SI{300}{\pc}$,
  class = symbol
}

\DeclareAcronym{velDispB}{
  short = \ensuremath{\sigma_*},
  long  = bulge velocity dispersion,
  class = symbol
}

\DeclareAcronym{velDisp-proj}{
  short = \ensuremath{\hat{\sigma}},
  long  = projected velocity dispersion,
  class = symbol
}

\DeclareAcronym{M-dm}{
  short = \ensuremath{M_{\rm DM}},
  long  = dark matter mass,
  class = symbol
}

\DeclareAcronym{v-max}{
  short = \ensuremath{v_{\rm max}},
  long  = maximal velocity,
  class = symbol
}

\DeclareAcronym{v-min}{
  short = \ensuremath{v_{\rm min}},
  long  = minimal velocity,
  class = symbol
}

\DeclareAcronym{v-obs}{
  short = \ensuremath{v_{\rm obs}},
  long  = observed total velocity,
  class = symbol
}

\DeclareAcronym{v-bar}{
  short = \ensuremath{v_{\rm bar}},
  long  = velocity of the baryonic component,
  class = symbol
}

\DeclareAcronym{v-gas}{
  short = \ensuremath{v_{\rm gas}},
  long  = velocity of the gas component,
  class = symbol
}

\DeclareAcronym{v-star}{
  short = \ensuremath{v_{\rm star}},
  long  = stellar velocity,
  class = symbol
}

\DeclareAcronym{v-dm}{
  short = v_{\rm DM},
  long  = velocity of the dark matter component,
  class = symbol
}

\DeclareAcronym{r-max}{
  short = \ensuremath{r_{\rm max}},
  long  = radius at the maximum circular velocity,
  class = symbol
}

\DeclareAcronym{r-max-d}{
  short = \ensuremath{r_{\rm max(d)}},
  long  = radius at the maximum circular velocity (dwarfs),
  class = symbol
}

\DeclareAcronym{rh-d}{
  short = \ensuremath{r_{h\rm(d)}},
  long  = one-halo length scale of the RAR model (dwarfs),
  class = symbol
}

\DeclareAcronym{Mh-d}{
  short = \ensuremath{M_{h\rm(d)}},
  long  = halo mass (dwarfs),
  class = symbol
}

\DeclareAcronym{r-max-s}{
  short = \ensuremath{r_{\rm max(s)}},
  long  = radius at the maximum circular velocity (spirals),
  class = symbol
}

\DeclareAcronym{rh-s}{
  short = \ensuremath{r_{h\rm(s)}},
  long  = one-halo length scale of the RAR model (spirals),
  class = symbol
}

\DeclareAcronym{Mh-s}{
  short = \ensuremath{M_{h\rm(s)}},
  long  = halo mass (spirals),
  class = symbol
}

\DeclareAcronym{r-max-e}{
  short = \ensuremath{r_{\rm max(e)}},
  long  = radius at the maximum circular velocity (ellipticals),
  class = symbol
}

\DeclareAcronym{rh-e}{
  short = \ensuremath{r_{h\rm(e)}},
  long  = one-halo length scale of the RAR model (ellipticals),
  class = symbol
}

\DeclareAcronym{Mh-e}{
  short = \ensuremath{M_{h\rm(e)}},
  long  = halo mass (ellipticals),
  class = symbol
}

\DeclareAcronym{r-max-bcg}{
  short = \ensuremath{r_{\rm max(bcg)}},
  long  = radius at the maximum circular velocity (BCGs),
  class = symbol
}

\DeclareAcronym{rh-bcg}{
  short = \ensuremath{r_{h\rm(bcg)}},
  long  = one-halo length scale of the RAR model (BCGs),
  class = symbol
}

\DeclareAcronym{Mh-bcg}{
  short = \ensuremath{M_{h\rm(bcg)}},
  long  = halo mass (BCGs),
  class = symbol
}

\DeclareAcronym{Mbh}{
  short = \ensuremath{M_{\rm BH}},
  long  = black hole mass,
  class = symbol
}

\DeclareAcronym{MB}{
  short = \ensuremath{M_B},
  long  = absolute magnitude,
  class = symbol
}

\DeclareAcronym{Sigma0D}{
  short = \ensuremath{\Sigma_{0\rm D}},
  long  = dark matter surface density,
  class = symbol
}

\DeclareAcronym{rho0D}{
  short = \ensuremath{\rho_{0\rm D}},
  long  = central dark matter halo density,
  class = symbol
}

\DeclareAcronym{r0}{
  short = \ensuremath{r_0},
  long  = one-halo-scale-length of the Burkert profile,
  class = symbol
}

\DeclareAcronym{beta0-crit}{
  short = \ensuremath{\beta_0^{\rm cr}},
  long  = critical central temperature parameter,
  class = symbol
}

\DeclareAcronym{beta0-max}{
  short = \ensuremath{\beta_0^{\rm max}},
  long  = maximal central temperature parameter,
  class = symbol
}

\DeclareAcronym{beta0-min}{
  short = \ensuremath{\beta_0^{\rm min}},
  long  = minimal central temperature parameter,
  class = symbol
}

\DeclareAcronym{rc}{
  short = \ensuremath{r_c},
  long  = core radius,
  class = symbol
}

\DeclareAcronym{rh}{
  short = \ensuremath{r_h},
  long  = halo radius,
  class = symbol
}

\DeclareAcronym{rb}{
  short = \ensuremath{r_{\rm b}},
  long  = boundary radius,
  class = symbol
}

\DeclareAcronym{rb-max}{
  short = \ensuremath{r_{\rm b}^{\rm max}},
  long  = boundary radius,
  class = symbol
}

\DeclareAcronym{Mc}{
  short = \ensuremath{M_c},
  long  = core mass,
  class = symbol
}

\DeclareAcronym{Mc-crit}{
  short = \ensuremath{M_c^{\rm cr}},
  long  = critical core mass,
  class = symbol
}

\DeclareAcronym{Mc-max}{
  short = \ensuremath{M_c^{\rm max}},
  long  = maximum core mass,
  class = symbol
}

\DeclareAcronym{Mc-min}{
  short = \ensuremath{M_c^{\rm min}},
  long  = minimum core mass,
  class = symbol
}

\DeclareAcronym{Mh}{
  short = \ensuremath{M_h},
  long  = halo mass,
  class = symbol
}

\DeclareAcronym{Mtot}{
  short = \ensuremath{M_{\rm tot}},
  long  = total dark matter mass,
  class = symbol
}

\DeclareAcronym{Mtot-max}{
  short = \ensuremath{M_{\rm tot}^{\rm max}},
  long  = maximum total dark matter mass,
  class = symbol
}

\DeclareAcronym{Mtot-min}{
  short = \ensuremath{M_{\rm tot}^{\rm min}},
  long  = minimum total dark matter mass,
  class = symbol
}

\DeclareAcronym{rhoc}{
  short = \ensuremath{\rho_c},
  long  = core density,
  class = symbol
}

\DeclareAcronym{rhop}{
  short = \ensuremath{\rho_{\rm pl}},
  long  = plateau density,
  class = symbol
}

\newcommand{\ValMcoreCR	}{\SI{2.2E8}{\Msun}}

\newcommand{\ValOrhD		}{\SI{400}{\parsec}}
\newcommand{\ValOMhD		}{\SI{3E7}{\Msun}}

\newcommand{\ValOrhS		}{\SI{50}{\kilo\parsec}}
\newcommand{\ValOMhS		}{\SI{1E12}{\Msun}}

\newcommand{\ValOrhE		}{\SI{90}{\kilo\parsec}}
\newcommand{\ValOMhE		}{\SI{5E12}{\Msun}}

\newcommand{\ValOrhBCG	}{\SI{600}{\kilo\parsec}}
\newcommand{\ValOMhBCG	}{\SI{3E14}{\Msun}}

\newcommand{\MBH 			}{M_{\rm BH}}

\newcommand{\Mtot 		}{M_{\rm tot}}

\newcommand{\SYMvel		}{V}
\newcommand{\SYMvbulge}{V_\mathrm{b}}
\newcommand{\SYMvdisk	}{V_\mathrm{d}}
\newcommand{\SYMvgas	}{V_\mathrm{g}}

\newcommand{\SYMvobs	}{V_\mathrm{obs}}
\newcommand{\SYMvbar	}{V_\mathrm{bar}}
\newcommand{\SYMvdark	}{V_\mathrm{DM}} 

\newcommand{\SYMacc		}{a}

\DeclareAcronym{DM}{
  short = DM,
  long  = Dark Matter,
  class = abbrev
}

\DeclareAcronym{CDM}{
  short = CDM,
  long  = Cold Dark Matter,
  class = abbrev-scheme
}

\DeclareAcronym{WDM}{
  short = WDM,
  long  = Warm Dark Matter,
  class = abbrev-scheme
}

\DeclareAcronym{LCDM}{
  short = $\Lambda$CDM,
  long  = $\Lambda$ Cold Dark Matter,
  class = abbrev-scheme
}

\DeclareAcronym{LWDM}{
  short = $\Lambda$WDM,
  long  = $\Lambda$ Warm Dark Matter,
  class = abbrev-scheme
}

\DeclareAcronym{RAR}{
  short = RAR,
  long  = Ruffini-Argüelles-Rueda,
  class = abbrev-model
}

\DeclareAcronym{NFW}{
  short = NFW,
  long  = Navarro-Frenk-White,
  class = abbrev-model
}

\DeclareAcronym{FDM}{
  short = FDM,
  long  = fuzzy DM,
  class = abbrev-model
}

\DeclareAcronym{BH}{
  short = BH ,
  long  = Black Hole ,
  class = abbrev-co
}

\DeclareAcronym{SMBH}{
  short = SMBH ,
  long  = supermassive Black Hole ,
  class = abbrev-co
}

\DeclareAcronym{IMBH}{
  short = IMBH,
  long  = intermediate-mass Black Hole,
  class = abbrev-co
}

\DeclareAcronym{SgrA}{
  short = SgrA*,
  long  = Sagittarius A*,
  class = abbrev-phen
}

\DeclareAcronym{GC}{
  short = GC,
  long  = Globular Cluster,
  class = abbrev-system
}

\begin{document}

\begin{frontmatter}
\title{Novel constraints on fermionic dark matter from galactic observables II:\\galaxy scaling relations}
	
\author[cct,icranet]{C.~R.~Argüelles}
\ead{carlos.arguelles@icranet.org}
\author[icranet,icra,nice]{A.~Krut}
\ead{andreas.krut@icranet.org}
\author[icranet,icra,cbpf]{J.~A.~Rueda}
\ead{jorge.rueda@icra.it}
\author[icranet,icra,cbpf]{R.~Ruffini}
\ead{ruffini@icra.it}

\address[cct]{Instituto de Astrofísica de La Plata (CCT La Plata, CONICET, UNLP), Paseo del Bosque, B1900FWA La Plata, Argentina}
\address[icranet]{ICRANet, Piazza della Repubblica 10, I--65122 Pescara, Italy}
\address[icra]{Dipartimento di Fisica and ICRA, Sapienza Università di Roma, P.le Aldo Moro 5, I--00185 Rome, Italy}
\address[nice]{University of Nice-Sophia Antipolis, 28 Av. de Valrose, 06103 Nice Cedex 2, France}
\address[cbpf]{ICRANet-Rio, CBPF, Rua Dr.~Xavier Sigaud 150, Rio de Janeiro, RJ, 22290--180, Brazil}

\begin{abstract}
We have recently introduced in paper I an extension of the Ruffini-Arg\"uelles-Rueda (RAR) model for the distribution of DM in galaxies, by including for escape of particle effects. Being built upon self-gravitating fermions at finite temperatures, the RAR solutions develop a characteristic \textit{dense quantum core-diluted halo} morphology which, for fermion masses in the range $mc^2 \approx 10-345$ keV, was shown to provide good fits to the Milky Way rotation curve. We study here for the first time the applicability of the extended RAR model to other structures from dwarfs to ellipticals to galaxy clusters, pointing out the relevant case of $mc^2 = 48$ keV. By making a full coverage of the remaining free parameters of the theory, and for each galactic structure, we present a complete family of astrophysical RAR profiles which satisfy realistic halo boundary conditions inferred from observations. Each family-set of RAR solutions predicts given windows of total halo masses and central quantum-core masses, the latter opening the interesting possibility to interpret them as alternatives either to intermediate-mass BHs (for dwarf galaxies), or to supermassive BHs (SMBHs, in the case of larger galaxy types). The model is shown to be in good agreement with different observationally inferred scaling relations such as: (1) the Ferrarese relation connecting DM halos with supermassive dark central objects; and (2) the nearly constant DM surface density of galaxies. Finally, the theory provides a natural mechanism for the formation of SMBHs of few $10^8 M_\odot$ via the gravitational collapse of unstable DM quantum-cores.
\end{abstract}


\end{frontmatter}

\section{Introduction}
\label{sec:introduction}
The problem of describing \ac{DM} halos in terms of fundamental particles has gained considerable attention in the last years, given they may provide solutions to many of the unsuccessful predictions of the \ac{CDM} paradigm arising below $\sim\SI{10}{\kilo\parsec}$ scales. The majority of such models are comprised within the following three different approaches:
(i) The case of ultra light bosons with masses in the range $m_bc^2\sim \SIrange{1}{100E-22}{\eV}$, known as ultra light \ac{DM}, \ac{FDM} or scalar field \ac{DM} \citep{1983PhLB..122..221B,1994PhRvD..50.3650S,2000PhRvL..85.1158H,2001PhRvD..63f3506M,2012MNRAS.422..282R,2017PhRvD..95d3541H,2018arXiv180500122B};
(ii) the case of Thomas-Fermi models based on fully-degenerate fermions with masses $m_{df}c^2\sim$ few $\SI{E2}{\eV}$ \citep{2013NewA...22...39D,2015JCAP...01..002D,2017MNRAS.467.1515R}.  We also include in this group similar models based on self-gravitating fermions but in the dilute regime (i.e. Boltzmannian-like) which, however, \textit{do not} imply an explicit particle mass dependence when contrasted with halo observables \citep[see e.g.][]{2014MNRAS.442.2717D};
(iii) the \ac{RAR} model based on semi-degenerate configurations of self-gravitating fermions accounting for finite temperature and for relativistic effects, with masses in the range $m_fc^2\sim$ few $\SIrange{E1}{E2}{\kilo\eV}$ \citep{2013pdmg.conf30204A,2014IJMPD..2342020A,2015MNRAS.451..622R,2015ARep...59..656S,2016JCAP...04..038A,2016PhRvD..94l3004G,2018PDU....21...82A}.

One of the main interesting aspects of the above models is the particle mass dependence on their density profiles (besides the other physically-motivated free parameters), differently to the case of phenomenological profiles existing in the literature aiming to fit results from classical N-body numerical simulations. Moreover, such kind of self-gravitating systems of particles opens the possibility to have direct access to the nature, mass, and explicit phase-space distributions dependence at the onset of halo formation (see e.g., ~\citealp{2017PhRvD..95d3541H} and refs. therein, for the case of bosons and, e.g.~\citealp{2018PDU....21...82A} and refs. therein, for the case of fermions). 

However, not all of the above particle-motivated \ac{DM} halo models have the required cosmological and astrophysical properties when contrasted with different observational data-sets. Some fundamental problems remain mainly within the cases (i) and (ii) such as: (a) galaxy-scaling relations; (b) Ly$\alpha$ forest constraints; (c) nearby disk galaxies with high resolution rotation curve features. 

Concerning the first point (a), we could mention the recent results in \citet{2018PhRvD..98b3513D} which demonstrated that standard \ac{FDM} models without self-interactions\footnote{
	Composed, for example, by axion-like particles with $m_bc^2\sim\SI{E-22}{\eV}$ as the one proposed in \citet{2000PhRvL..85.1158H,2017PhRvD..95d3541H}.
} are ruled out when asking to follow the \ac{DM}-surface-density Universal relation \citep{2009MNRAS.397.1169D,2017MNRAS.470.2410R}. That is, while this observational relation imposes that the central \ac{DM} halo density $\rho_c$ scales with the inverse of the core radius $r_c$ to some power $\beta\approx 1$ (i.e. $\rho_c\sim r_c^{-1}$); the \ac{FDM} theory yields an inverse proportionality relation but with a quite different power $\beta_{\rm FDM}=4$.\footnote{
	Alternative \ac{FDM} models including for quartic-like self-interactions were also ruled out when contrasted against the above observed \ac{DM} halo Universal-relation in \citet{2018PhRvD..98b3513D}; see however \citet{2012MNRAS.422..282R} for other kind of self-interacting bosons.
}

On a different observational ground, and moving to the issue (b), the \ac{DM} particle mass constraints arising from recent Ly$\alpha$ forest observations, put severe bounds both to \ac{FDM} particles (i), as well as to fermions (ii)-(iii). The key concept to understand how all such bounds arise is to recognize that the Ly$\alpha$ forest can be explained in terms of the standard cosmological model \citep[see e.g.][]{1994ApJ...437L...9C}, and therefore it can be directly linked to the matter power spectrum by offering a complementary probe for it~\citep{1999ApJ...516..519H}. In the case of \ac{WDM} (exhibiting a linear matter power spectrum with a clear drop below a given threshold scale respect to the CDM paradigm), a lower bound of $\sim$ few \si{\kilo\eV} was found in \citet[][see also~\citealp{2017JCAP...06..047Y} for more general lower bounds above \si{\kilo\eV} including for sterile neutrino \ac{WDM}]{2013PhRvD..88d3502V} from Ly$\alpha$ constraints. Such bounds are in strong tension with the below-\si{\kilo\eV} Thomas-Fermi models cited above in (ii) as also discussed in \citet{2017MNRAS.467.1515R}, while it remains in agreement with the \ac{RAR} model introduced in (iii). Further discussion on \ac{FDM} models regarding issues (b) and (c) will be given below in \cref{sec:conclusion}.

We will show that the results of this work, together with the ones of the first article of this series~\citep[][hereafter Paper I]{2018PDU....21...82A} on the fit to the Milky Way full rotation curve, and the discussion of the constraints put to \acs{LCDM} and \acs{LWDM} cosmologies by the Ly$\alpha$ forest, favour instead the $\si{\kilo\eV}$ fermionic \ac{RAR} model (iii).

In next section we introduce the main features of the RAR model (as originally given in Paper I). In \cref{sec:constraints} we provide realistic halo boundary conditions as inferred from observational data coming from dwarf to ellitpical galaxies. In \cref{sec:results} we present the application of the extended \ac{RAR} theory to different galaxy types as well as galaxy clusters, and for fermion masses in the range limited by the Milky Way constraints. A more detailed description of the obtained families focusing on the configuration parameters is given in \cref{sec:analysis}. We also briefly compare in \cref{sec:robustness} the applied method with an alternative set of constraints for dwarfs to verify its robustness.

In addition, we demonstrate in \cref{sec:correlations} that the results of our model are consistent with (a) the observationally inferred correlations between the mass of the dark central object and total \ac{DM} halo mass \citep{2002ApJ...578...90F,2011Natur.469..377K,2015ApJ...800..124B}, and with (b) the observationally inferred (universal) value (within the $3-\sigma$ error bars) of the inner surface density of \ac{DM} halos \citep{2009MNRAS.397.1169D}, from dwarf to elliptical galaxies. Further \textit{predicting} for possible deviations of a constant surface density (i.e. slight rising trend or positive tilt in a linear fit) towards the brightest objects, in striking similitude with the phenomenological results reported in \citet{2009arXiv0911.1774B}.

Finally, in \cref{sec:conclusion} we summarize the conclusions of our results.
\section{The RAR model}
\label{sec:RAR-model}

In Paper I \citep{2018PDU....21...82A} we have introduced an extension of the original RAR model \citep{2015MNRAS.451..622R}, by considering an spherical system of self-gravitating fermions presenting a cutoff in the Fermi-Dirac phase-space distribution. Such a (coarse-grained) phase-space distribution can be obtained as a (quasi) stationary solution of a generalized Fokker-Planck equation for fermions (including the physics of violent relaxation and escape of particles), proper to deal with non-linear DM halo formation \citep{2004PhyA..332...89C}; which is written below: 

\begin{equation}
f_c(\epsilon\leq\epsilon_c) = \frac{1-e^{(\epsilon-\epsilon_c)/kT}}{e^{(\epsilon-\mu)/kT}+1}, \qquad f_c(\epsilon>\epsilon_c)=
0\, ,
\label{fcDF}
\end{equation}
where $\epsilon=\sqrt{c^2 p^2+m^2 c^4}-mc^2$ is the particle kinetic energy, $\mu$ is the chemical potential with the particle rest-energy subtracted off, $T$ is the temperature, $k$ is the Boltzmann constant, $h$ is the Planck constant, $c$ is the speed of light, and $m$ is the fermion mass. We do not include the presence of anti-fermions, i.e.~we consider temperatures $T \ll m c^2/k$. The full set of (functional) dimensionless-parameters of the model are defined by the temperature, degeneracy and cutoff parameters, $\beta=k T/(m c^2)$, $\theta=\mu/(k T)$ and $W=\epsilon_c/(k T)$, respectively.

The corresponding 4-parametric fermionic equation of state (at given radius $r$): $\rho(\beta,\theta,W,m), P(\beta,\theta,W,m)$, is directly obtained as the corresponding integrals (bounded from above by $\epsilon \leq \epsilon_c$)  over momentum space of $f_c(p)$, given in equations ($1$) and ($2$) in Paper I. Such components of the equation of state corresponds to the diagonal part of the stress-energy tensor in the Einstein equations, which are solved under the perfect fluid approximation within a background metric with spherical symmetry ${\rm d}s^2 = e^{\nu}c^2 {\rm d}t^2 -e^{\lambda}{\rm d}r^2 -r^2 {\rm d}\Theta^2 -r^2\sin^2\Theta {\rm d}\phi^2$ (with ($r$,$\Theta$,$\phi$) the spherical coordinates, and $\nu$ and $\lambda$ only depending on the radial coordinate $r$). The system of Einstein equations is solved together with the Tolman and Klein thermodynamic equilibrium conditions, and (particle) energy conservation along a geodesic as given in equations $5,6$ and $7$ of Paper I respectively. Finally, the dimensionless system of non-linear ordinary integro-differential equations reads:

\begin{align}
	\frac{{\rm d}\hat M_{\rm DM}}{{\rm d}\hat r}&=4\pi\hat r^2\hat\rho, \label{eq:eqs1}\\
	\frac{{\rm d}\theta}{{\rm d}\hat r}&=-\frac{1-\beta_0(\theta-\theta_0)}{\beta_0}
    \frac{\hat M_{\rm DM}+4\pi\hat P\hat r^3}{\hat r^2(1-2\hat M_{\rm DM}/\hat r)},\label{eq:eqs2}\\
    \frac{{\rm d}\nu}{{\rm d}\hat r}&=\frac{2(\hat M_{\rm DM}+4\pi\hat P\hat r^3)}{\hat r^2(1-2\hat M_{\rm DM}/\hat r)}, \\
    \beta(\hat r)&=\beta_0 e^{\frac{\nu_0-\nu(\hat r)}{2}}, \\
    W(\hat r)&=W_0+\theta(\hat r)-\theta_0\, .\label{eq:Cutoff}
\end{align}
Such that in the limit $W\to\infty$ (i.e. no particle escape: $\epsilon_c\to\infty$) these system reduce to the equations considered in the original RAR model \citep{2015MNRAS.451..622R}. We have introduced the same dimensionless quantities as in the original RAR model formulation: $\hat r=r/\chi$, $\hat M_{\rm DM}=G M_{\rm DM}/(c^2\chi)$, $\hat\rho=G \chi^2\rho/c^2$, $\hat P=G \chi^2 P/c^4$, where $\chi=2\pi^{3/2}(\hbar/mc)(m_p/m)$ and $m_p=\sqrt{\hbar c/G}$ the Planck mass. We note that the constants of the Tolman and Klein conditions are evaluated at the center $r=0$, indicated with a subscript `0'.

The task of Paper I (for the Milky Way), was thus to solve the system (\ref{eq:eqs1}--\ref{eq:Cutoff}), for given regular initial conditions at the center, $[M_{\rm DM}(0)=0,\theta(0)=\theta_0,\beta(0)=\beta_0,\nu(0)=0,W(0)=W_0]$, for different DM particle mass $m$, to find solutions consistent with well-constrained DM halo observables of the Galaxy (see also beginning of next section). One of the main features of the RAR solutions is the fact that they are solved at finite temperature for \textit{positive} central degeneracies $\theta_0>0$ (instead of the known diluted Boltzmannian-like regime $\theta_0<0$), giving rise to a degenerate core at the center of the halo which always fulfill the quantum condition: $\lambda_B\gtrsim 3 l_c$ (where $l_c\sim n_c^{-1/3}$ is the interparticle mean distance within the core (with $n_c$ the core particle density) and $\lambda_B=h/(2\pi m k T_c)^{1/2}$ the thermal de-Broglie wavelength at the core). The dense and degenerate core is followed by a transition from positive to negative values of $\theta$ where quantum corrections are still important, finally reaching the region of highly negative values corresponding to a Boltzmannian diluted regime (see fig. ($2$) in Paper I). Such a core-halo transition with the different physical behaviours in $\theta$ is easily understood by the fact that the fermions are immerse in an external gravitational field, leading to a radial gradient (monotonically decreasing) of the degeneracy. 

This core-halo behaviour in the intrinsic RAR model parameters is reflected in the DM density and rotation curve profiles as follows (whose solution for the Milky Way DM halo is displayed in figs. ($3$) and ($4$) in Paper I, for different RAR model paramters):
\begin{enumerate}
\item 
an inner core with radius $r_c$ of almost constant density governed by quantum degeneracy (see the region of high positive values of the degeneracy parameter in upper panel of fig. 2 of Paper I);
\item
an intermediate region with a sharply decreasing density distribution followed by an extended plateau, where quantum corrections are still important (see the region of transition from positive to negative values of degeneracy in upper panel of fig. 2 of Paper I); and 
\item
a Boltzmannian density tail (see highly negative values of the degeneracy parameter in upper panel of fig. 2 in Paper I) showing a behavior $\rho\propto r^{-n}$ with $n>2$ due to the cutoff constraint (when $W_0$ approaches 0), as can be seen from lower panel of fig. 2 in Paper I. 
\end{enumerate}

\section{Observational constraints}
\label{sec:constraints}
In the case of our Galaxy, thanks to the vast amount of rotation curve data, i.e. as obtained in \citealp{2013PASJ...65..118S}, from inner bulge to outer halo, we were able to identify three relevant \textit{observables} as the boundary conditions to be imposed to the RAR system of differential equations (\ref{eq:eqs1}--\ref{eq:Cutoff}): a dark core mass $\acs{M-dm}(r=\acs{rc}) \equiv \acs{Mc}=\SI{4.2E6}{\Msun}$ (alternative to the central BH) and two well constrained dark halo masses: $\acs{M-dm}(r=\SI{12}{\kilo\parsec}) = \SI{5E10}{\Msun}$; $\acs{M-dm}(r=\SI{40}{\kilo\parsec})=\SI{2E11}{\Msun}$ (see Paper I: \citealp{2018PDU....21...82A}). With such three observationally well constrained mass values for the Milky Way it was possible to obtain in Paper I the three (free) RAR model parameters ($\beta_0$,$\theta_0$,$W_0$) for different particle masses $m$ in the $\si{\kilo\eV}$ range.

It is now natural to ask whether or not the \ac{RAR} model can explain the observational properties of other types of galaxies or even galaxy clusters, in the same range of \ac{DM} particle mass obtained from the Milky way analysis. We therefore proceed to show how, for a fixed particle mass $m$, our model leads to an extensive three-parametric ($\theta_0,\beta_0,W_0$) family of dark halos with parameters ranging from the ones of dwarf, to the ones in elliptical galaxies extending until galaxy clusters, harboring at the same time a semi-degenerate quantum core at each center.

When dealing with different galactic structures, located far away from us, the observational inferences of the \ac{DM} content are limited to a narrow window of galaxy radii, usually lying just above the baryonic dominance region (i.e.~typically up to several half-light radii). Generally, there is no observational access neither for the possible detection of a dark compact object at the centre nor for constraining the boundary of the \ac{DM} halo at the virial radius scale. This is contrary to the case of the Milky Way, thus requiring a different phenomenology which is described in next.

We adopt here a similar methodology compared to the Milky Way analysis \citep{2018PDU....21...82A}, but limited to radial halo extents where observational data is available, allowing to constraint the \ac{DM} halo mass either in a model independent (or dependent) manner. In particular we will select as the only boundary conditions taken from observables a characteristic halo radius \acs{rh} with the corresponding halo mass $\acs{Mh} \equiv \acs{M-dm}(\acs{rh})$. The halo radius is defined as the location of the maximum of the halo rotation curve which we adopt as the one-halo scale length of our model. Thus, we define in next the parameters adopted for the different \ac{DM} halos as constrained from observations in typical dwarf spheroidal (dSph), spiral, elliptical to galaxy cluster structures.

\subsection{Typical dSph galaxies}
We consider the eight best resolved dwarf satellites of the Milky Way as analyzed in \citet{2009ApJ...704.1274W} by solving the Jeans equations, using large (stellar) kinematic data sets and including for orbital anisotropy.

There, it was reported a \ac{DM} model-independent evidence of a maximum circular velocity (\acs{v-max}) in the \ac{DM} halo of the Fornax dwarf (see fig.~2 in \citealp{2009ApJ...704.1274W}). Such an evidence was found by comparing the theoretical projected dispersion velocity (from Jeans equation) with the observed one (through a Markov-Chain Monte-Carlo method), using a 4-parametric generalized Hernquist mass model for the halo, the latter allowing either for cored or cuspy density profiles depending on the free parameters. The best fit to \acs{v-max} was found independently of the couple of free parameters which control the \ac{DM} shapes, i.e. in a \ac{DM} independent way.

In the other seven cases, a \ac{DM} model-dependent evidence for a circular velocity peak was found assuming either cuspy (e.g. \acs{NFW}) or cored (e.g. cored-Hernquist) \ac{DM} halos density profiles. 

In all eight cases the inferred radii and masses at the maximum circular velocity (supported by data) are $\acs{r-max-d} \sim$~few $\SI{E2}{\parsec}$ and $M(\acs{r-max-d}) \sim$~few $\SI{E7}{\Msun}$. These values have been obtained by assuming a cored-Hernquist \ac{DM} profile,\footnote{
	Somewhat larger values $\acs{r-max-d} \sim \SI{1}{\kilo\parsec}$ and $M(\acs{r-max-d}) \sim \SI{E8}{\Msun}$ are obtained for cuspy profiles. Though, the latter are disfavored respect to cored ones for dSph, as recently reviewed in \citet{2017ARA&A..55..343B}.
} similar to the \ac{RAR} profiles here presented (see e.g. \citealp{2015MNRAS.451..622R,2016JCAP...04..038A} for the \ac{RAR} halo fits to isothermal and Burkert profiles respectively).

The radius of maximal circular velocity we identify as the one-halo length scale of the RAR model $\acs{rh-d} \equiv \acs{r-max-d}$ with the corresponding \ac{DM} halo mass $\acs{Mh-d} \equiv M(\acs{r-max-d})$. Thus, as allowed by data, we adopt throughout this work the following fiducial values for the characteristic \ac{DM} halo properties for typical dSphs:
\begin{align}
\label{eqn:constraint:dwarf:Mh}
\acs{rh-d} &= \ValOrhD \\
\label{eqn:constraint:dwarf:rh}
\acs{Mh-d} &= \ValOMhD
\end{align}

\subsection{Typical spiral galaxies}
We consider some nearby disk galaxies observed in high resolution from the THINGS data sample \citep{2008AJ....136.2648D}, where \ac{DM} model-independent evidence for a maximum in the halo rotation curves is provided. Such an evidence is obtained by accounting for baryonic (stars and gas) components --- thanks to the inclusion of infrared data from the \textit{Spitzer} telescope --- in addition to the (total) observed rotation curve from the HI tracers. They calculated along the full observed data coverage, the \ac{DM} contribution $v_{\rm halo}$ to the observed rotation curve $v_{\rm obs}$, by means of $v^2_{\rm halo}=v^2_{\rm obs} - v^2_{\rm bar}$, through the corresponding build up of mass models for the baryonic components $v^2_{\rm bar}=\Upsilon_*v^2_{*}+v^2_{\rm gas}$ (with $\Upsilon_*$ the stellar mass-to-light ratio).

This analysis shows galaxies with extended enough data coverage (mainly corresponding to the larger and more luminous) supporting for evidence of a maximum in the circular velocity (see gray curves within fig.~63 in \citealp{2008AJ....136.2648D}). The maximum values for radii and velocity in the more luminous galaxies ($\acs{MB} \lesssim -20$) are expected to be bounded from above and below as, $\acs{r-max-s} \approx \SIrange{10}{80}{\kilo\parsec}$, and $v(\acs{r-max-s}) \approx \SIrange{70}{310}{\kilo\metre/\second}$, further implying $M(\acs{r-max-s}) \approx \SIrange{E10}{2E12}{\Msun}$. The bounds for $\acs{r-max-s}$ and $v(\acs{r-max-s})$ are reported in \citet[fig.~63]{2008AJ....136.2648D} using NFW models with data supporting up to $\approx \SI{50}{\kilo\parsec}$. Note the similar behaviour between the NFW and the \ac{RAR} models on the halo scales of interest, as shown in \citet[fig.~3]{2018PDU....21...82A} for the case of the Milky Way.

Analogue to dwarf galaxies we identify the radius of the maximal circular velocity as the one-halo length scale $\acs{rh-s} \equiv \acs{r-max-s}$ with the corresponding \ac{DM} halo mass $\acs{Mh-s} \equiv M(\acs{r-max-s})$. Thus, as allowed by data, we adopt throughout this work the following fiducial values for the characteristic \ac{DM} halo properties for typical spirals:
\begin{align}
\label{eqn:constraint:spiral:Mh}
\acs{rh-s} &= \ValOrhS \\
\label{eqn:constraint:spiral:rh}
\acs{Mh-s} &= \ValOMhS
\end{align}

\subsection{Typical elliptical galaxies}
We consider a sample of elliptical galaxies from \citet{2005ApJ...635...73H}, studied via weak lensing signals, and further analyzed in \citet{2009MNRAS.397.1169D} by providing halo mass models for the tangential shear of the distorted images (i.e. galaxies taken from Appendix A in \citealp{2009MNRAS.397.1169D}).

We also consider the iconic case of the largest and closest elliptical M87 as studied in \citet{2001ApJ...553..722R}, accounting for combined halo mass tracers such as stars, \acp{GC} and X-ray data (see also \citealp{1995MNRAS.274.1093N}).

Kinematic measurements (e.g.~\acp{GC}) can probe distances up to several $\SI{E1}{\kilo\parsec}$, while X-ray and weak lensing data can reach much further distances up to several $\SI{E2}{\kilo\parsec}$. Thus, the latter usually allow for (\ac{DM} model-dependent) evidence of a maximum circular velocity on halo scales, where data supports. Such an evidence was provided in \citet{2009MNRAS.397.1169D} and \citet{2001ApJ...553..722R} through the \ac{DM} profiles (i.e. Burkert and \acs{NFW} respectively), to obtain best fits to the full data coverage in the galaxies there considered. Providing the following maxima values: $\acs{r-max-e} \approx \SI{100}{\kilo\parsec}$ and $M(\acs{r-max-e}) \approx \SI{E12}{\Msun}$ (in the case of the more luminous ellipticals with $M_{B}<-20$, following Burkert) up to $\approx \SI{E13}{\Msun}$ (in the case of M87, following \acs{NFW}).

Again, we identify here the radius of the maximal circular velocity as the one-halo length scale $\acs{rh-e} \equiv \acs{r-max-e}$ with the corresponding \ac{DM} halo mass $\acs{Mh-e} \equiv M(\acs{r-max-e})$. Thus, as allowed by data, we adopt throughout this work the following fiducial values for the characteristic \ac{DM} halo properties for typical ellipticals:
\begin{align}
\label{eqn:constraint:elliptical:Mh}
\acs{rh-e} &= \ValOrhE \\
\label{eqn:constraint:elliptical:rh}
\acs{Mh-e} &= \ValOMhE
\end{align}

\subsection{Typical galaxy clusters}
We consider a sample of 7 brightest cluster galaxies (BCGs) from \citet{2013ApJ...765...25N}. In that work the luminous and dark components were separated to obtain DM distributions which can be well described by a generalized NFW (gNFW) model \citep{1996MNRAS.278..488Z}, developing a maximal velocity at the one-halo length scale $\acs{r-max-bcg}$. Such morphology is shown in \citet{2013ApJ...765...25N} to be supported by data sets including for weak lensing and stellar kinematics, covering a radial extend from $\SI{10}{\kilo\parsec}$ up to $\SI{3}{\mega\parsec}$. In all cases the halo radii and masses can be inferred from such gNFW profiles and read $\acs{r-max-bcg} \approx \SIrange{E2}{E3}{\kilo\parsec}$, and $M(\acs{r-max-bcg}) \approx \SIrange{E14}{E15}{\Msun}$ respectivley.

Equivalent to other galaxy types, we identify here the radius of the maximal circular velocity as the one-halo length scale $\acs{rh-bcg} \equiv \acs{r-max-bcg}$ with the corresponding \ac{DM} halo mass $\acs{Mh-bcg} \equiv M(\acs{r-max-bcg})$. For simplicity we take the mean values of the inferred radii and masses as typical values for BCGs. Thus, as allowed by data, we adopt throughout this work the following fiducial values for the characteristic \ac{DM} halo properties for typical BCGs:
\begin{align}
\label{eqn:constraint:BCG:Mh}
\acs{rh-bcg} &= \ValOrhBCG \\
\label{eqn:constraint:BCG:rh}
\acs{Mh-bcg} &= \ValOMhBCG
\end{align}

\subsection{Method}
\label{sec:other:method}
The halo values adopted above for each (representative) galaxy or BCG, are such that 
\begin{itemize}
\item they correspond to \ac{DM} dominated halos as carefully checked in each observational work cited above;
\item they do not account for the (total) virial \ac{DM} mass due to natural observational limitations, but they rather represent the \ac{DM} halo characteristics somewhat outside the region of baryon dominance.
\end{itemize}

We thus systematically calculate (for the relevant example of $mc^2=\SI{48}{\kilo\eV}$ as motivated by the Milky Way analysis in Paper I) the RAR solutions represented through the configuration parameter ($\beta_0, \theta_0, W_0$), which match the halo constraints \acs{rh} and \acs{Mh} with a tolerance $\tau = \num{E-3}$ under the least-square condition

\begin{equation}
\sum \limits_{i=1}^N \abs{1 - \frac{f_i^{\rm RAR}(\vec b)}{y_i}}^2 < \tau^2.
\end{equation} 
Here, the observables $y_i$ for each galaxy or BCG case are compared with the predictions $f_i^{\rm RAR}(\vec b)$ for a parameter vector $\vec b = (\beta_0, \theta_0, W_0, m)$. The associated set of constraints $(\acs{rh},\acs{Mh})$ with $N=2$ are given in \cref{eqn:constraint:dwarf:Mh,eqn:constraint:dwarf:rh} for dwarfs, \cref{eqn:constraint:spiral:Mh,eqn:constraint:spiral:rh} for spirals, \cref{eqn:constraint:elliptical:Mh,eqn:constraint:elliptical:rh} for ellipticals and \cref{eqn:constraint:BCG:Mh,eqn:constraint:BCG:rh} for BCGs.

Notice that the observational constraints necessarily imply astrophysical RAR solutions which develop a maximum in the halo rotation curve (i.e. as set by $r_h$). Additionally, we request to the solutions one extra (underlying) physical condition, hereafter the \textit{quantum core} condition: the compact-core is stable or non-critical (i.e. it does not have the critical mass of gravitational collapse to a BH), fulfilling the quantum-statistics relation $\lambda_B\gtrsim 3 l_c$ in the core. These conditions define the full window of astrophysically allowed \ac{RAR}-family solutions. Importantly, the two halo constraints (for given $m$) provide a one-parametric family within the full ranges of the three configuration parameters ($\beta_0$, $\theta_0$, and $W_0$). Thus, the obtained values lay along a one-dimensional curve in the free configuration space and are limited from below and above; see \cref{sec:analysis} for further details.

\section{Results}
\label{sec:results}
The \ac{RAR} model provides, for each galaxy type and BCG with given \textit{observables} $(\acs{rh},\acs{Mh}$), a continuous set of solutions which is illustrated as a blue shaded region in \cref{fig:profiles}. In particular, we show five benchmark solutions, labeled with their central densities (from black to magenta, roughly enveloping the blue shaded regions), for \ac{DM} mass distributions $\acs{M-dm}(r)$, rotation curves $\acs{v-dm}(r)$ and density profiles $\rho(r)$. All solutions have been calculated for the relevant example of $mc^2=\SI{48}{\kilo\eV}$, with corresponding full set of free \ac{RAR} parameters as detailed in \cref{sec:analysis}.

They encompass a window of possible core and total masses, labeled as \acs{Mc} and \acs{Mtot}, for each galactic structure (see \cref{tbl:profiles} for benchmark numerical values and \cref{fig:ParameterAnalysis} for full range). Importantly, those mass windows are bounded from above and below as dictated by the astrophysical (i.e. \acs{v-max} at \acs{rh}) and \textit{quantum core} conditions (see \cref{sec:analysis} for details).

\def\ROOTPATH{figures/profiles}\begin{table*}
	\includegraphics[width=\hsize]{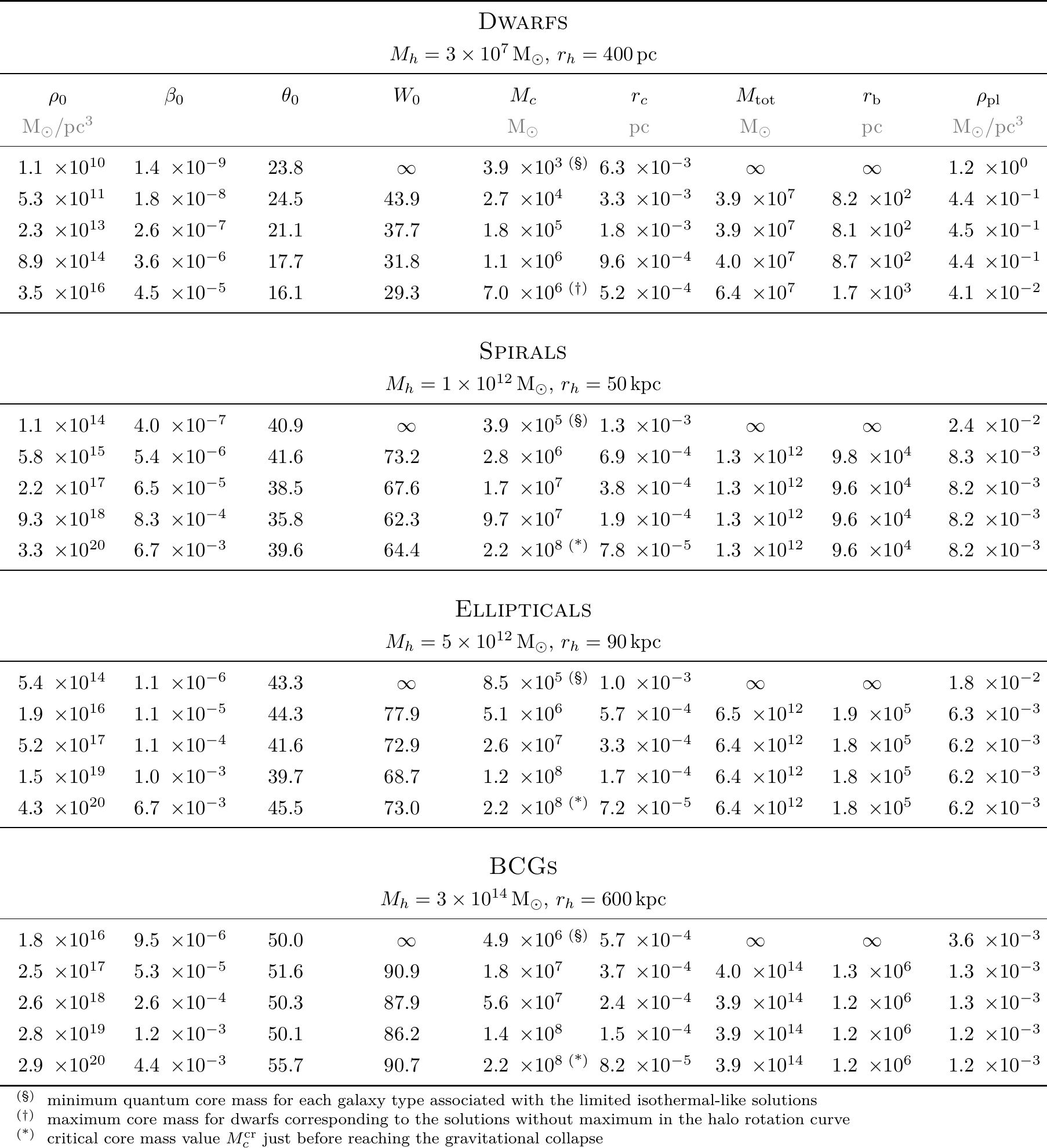}
	\caption{Free \ac{RAR} model parameters for the $5$ benchmark sets of \ac{DM} profiles given in \cref{fig:profiles}, for the case of $mc^2=\SI{48}{\kilo\eV}$, together with the DM masses (\acs{Mc}, \acs{Mtot}) and radii (\acs{rc}, \acs{rb}), as well as central and plateau densities ($\rho_0$, \acs{rhop}) for galactic structures from dwarfs to BCGs.}
	\label{tbl:profiles}
\end{table*}
\def\ROOTPATH{figures/profiles}\begin{figure*}%
	\centering%
	\includegraphics[width=\hsize]{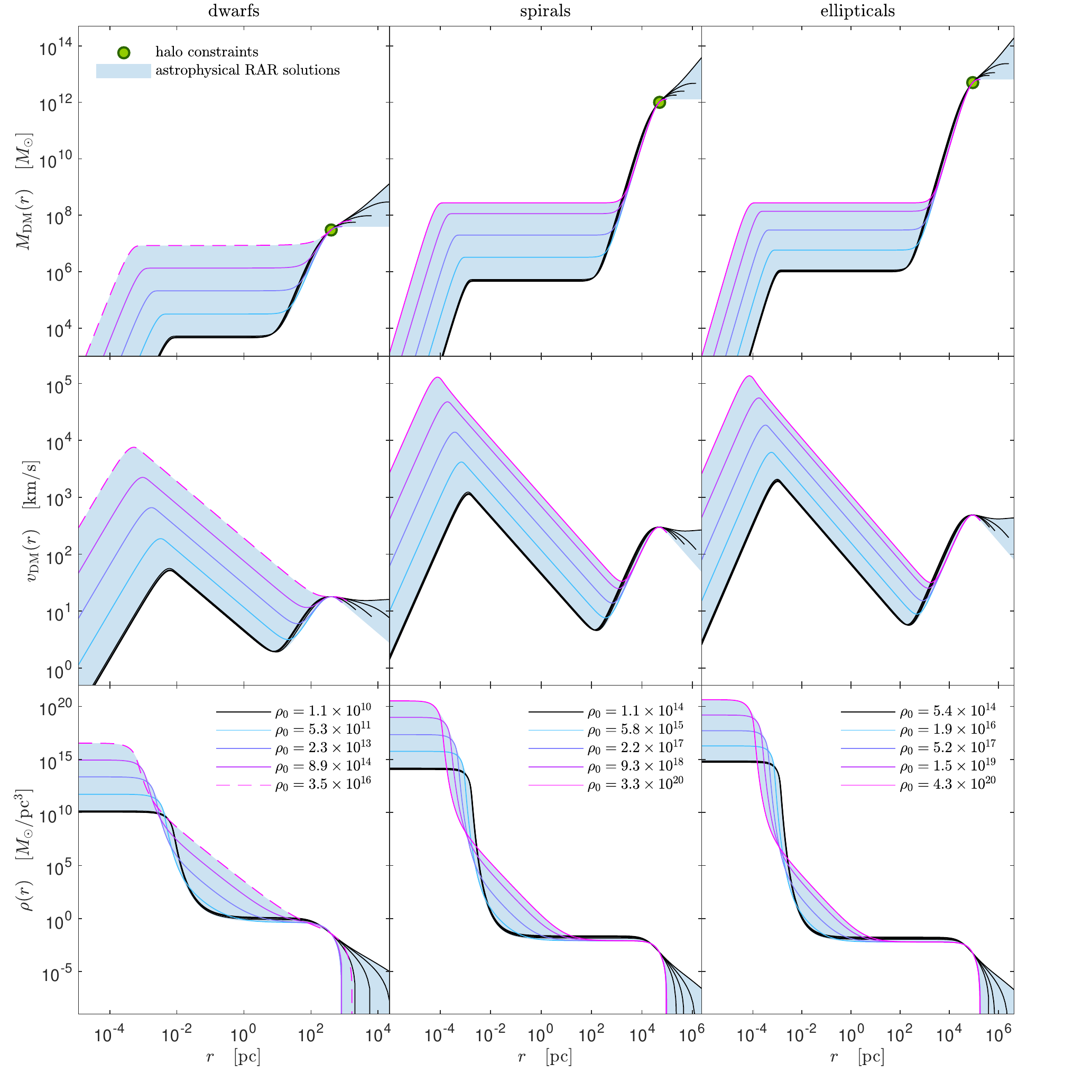}
	\caption{Astrophysical \ac{RAR} solutions, for the relevant case of $mc^2=\SI{48}{\kilo\eV}$, fulfill observationally given DM halo restrictions $(\acs{rh},\acs{Mh})$ for typical dwarf (left), spiral (middle) and elliptical galaxies (right). Shown are density profiles (bottom), rotation curves (middle) and DM mass distributions (top). The full window for each galaxy type is illustrated by a blue shaded region and enveloped approx. by $5$ benchmark solutions inside. Each solution is labeled with the central density in units of $\si{\Msun\parsec^{-3}}$ (the reader is referred to \cref{sec:analysis} for the explicit relation with the free parameters of the theory). The continuous-magenta curves, occurring only for spiral and elliptical galaxies, indicates the critical solutions which develop compact critical cores (before collapsing to a BH) of $\acs{Mc-crit} = \ValMcoreCR$. The dashed-magenta curves for dwarfs are limited (instead) by the astrophysical necessity of a maximum in the halo rotation curve. The bounding black solutions correspond to the ones having the minimum core mass (or minimum $\rho_0$) which in turn imply larger cutoff parameters (implying $\rho\propto r^{-2}$ when $W_0 \to\infty$).
	}
	\label{fig:profiles}
\end{figure*}

\begin{figure}[t!]
	\ContinuedFloat
	\captionsetup{list=off,format=cont}
	\centering%
	\includegraphics[width=0.8\hsize]{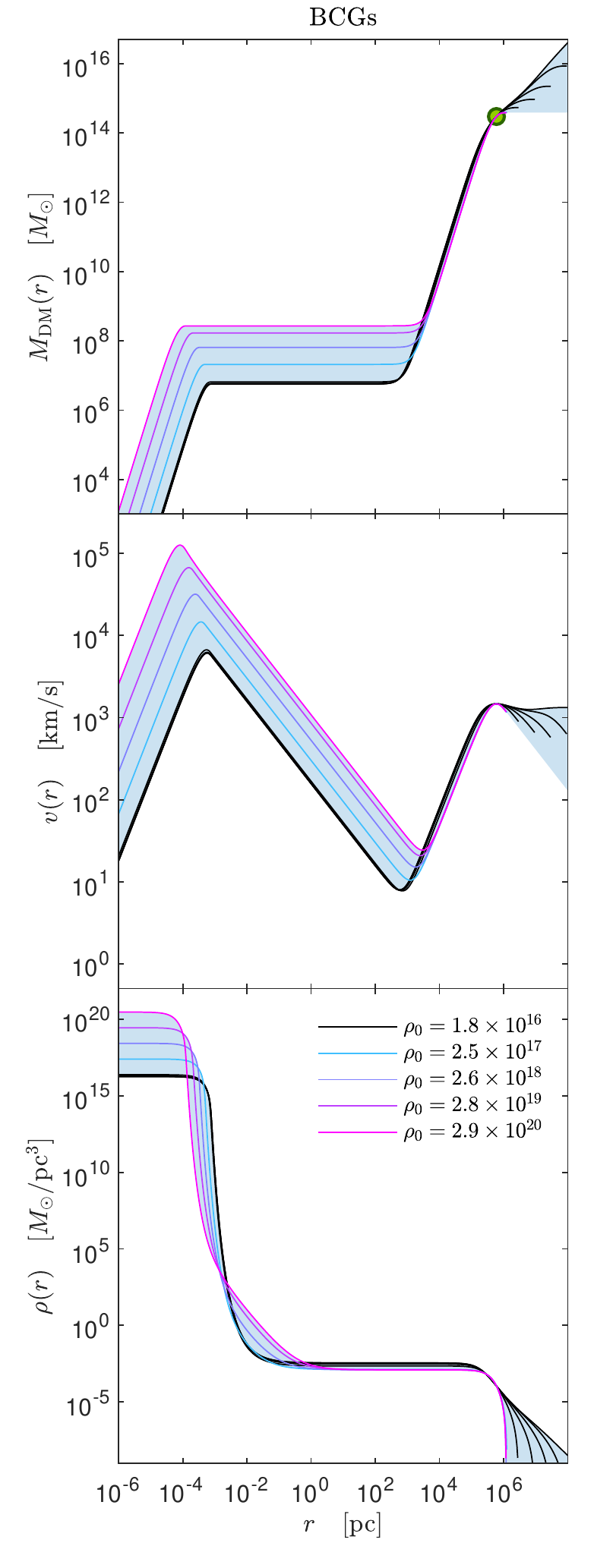}
	\caption{Profiles of BCGs.}
\end{figure}


The core mass $\acs{Mc} \equiv \acs{M-dm}(\acs{rc})$ is given at the core radius, defined at the first maximum of the rotation curve (analogously to the halo radius definition). The outermost \ac{DM} halo mass $\Mtot \equiv \acs{M-dm}(\acs{rb})$ is obtained at the border radius \acs{rb}, which is defined by $\rho(\acs{rb}) = 0$, above which there is no remaining particle energy as set by the \ac{RAR} cutoff parameter condition $W(\acs{rb}) = 0$ (see \cref{sec:RAR-model}). Both masses are an outcome of the \ac{RAR} family of astrophysical solutions and can be considered as a prediction of our theory (see \cref{tbl:profiles}).

The critical solutions, reaching the limiting core mass $\acs{Mc-crit} \approx \ValMcoreCR$ achieved only by the typical spiral and elliptical galaxies as well as for larger BCGs, are represented by the continuous magenta curves. The dashed magenta curve, in contrast, indicates the limiting (non-critical) solutions for typical dwarfs, where no maximum halo rotation curve is present (i.e. halo-scale \acs{v-max} and plateau-scale \acs{v-min} merge to a saddle point in the rotation curve).

On the other hand, the black curves correspond to the solutions acquiring the lowest possible central density $\rho_0$ but with a cutoff affecting the outer halo tails. These solutions develop more and more extended density tails resembling isothermal-like solutions, corresponding to $\rho\propto r^{-2}$ at large radii. Accordingly, the limiting case $W_0\to\infty$ resembles fully isothermal solutions, infinite in mass and size, in agreement with what was obtained in the original version of the model \citep{2015MNRAS.451..622R}.

The astrophysical conditions imposed to the solutions put no limit on the maximum value of $W_0$. Unless no other observational constraint is available (for a given galaxy) regarding the ending trend in the shape of the density tail, $W_0$ can increase indefinitely. Indeed, the larger $W_0$ the lesser the hardness in the falling-down shape of density profiles beyond \acs{rh} and the larger the boundary radius \acs{rb}. Of course, at some point \acs{rb} (and consequently \acs{Mtot}) will be excessively large to represent any reliable astrophysical halo. Therefore, those \ac{RAR} solutions must be discarded as physical ones (see \cref{sec:correlations} for the usage of a quantitative condition). We show, for completeness, at bottom left of \cref{fig:profiles} (black lines) the full plethora of density tails, corresponding with the specific minimum core mass solutions for each galaxy type.

It is important to make explicit that in all the cases analyzed the quantum core condition $\acs{lambda-B} \gtrsim 3 l_c$ is fulfilled. We obtain $\acs{lambda-B} \gtrsim 2.7\,l_c$ for dSphs, $\acs{lambda-B} \gtrsim 3.6\, l_c$  for typical spirals, $\acs{lambda-B} \gtrsim 3.8\, l_c$ for normal elliptical galaxies and $\acs{lambda-B} \gtrsim 4\, l_c$ for typical BCGs.

In sum we find that dark halos from dSph all the way up to BCG can be explained with regular and continuous distributions of the same type of fermions, having a particle mass of $mc^2\sim \SI{50}{\kilo\eV}$. Instead of massive BHs at their centres, our solutions develop massive and compact quantum cores with masses in the range 
\begin{itemize}
	\item $\acs{Mc} \approx \SIrange{3.9E3}{7.0E6}{\Msun}$ for typical dSphs
	\item $\acs{Mc} \approx \SIrange{3.9E5}{2.2E8}{\Msun}$ for typical spirals
	\item $\acs{Mc} \approx \SIrange{8.5E5}{2.2E8}{\Msun}$ for typical ellipticals
	\item $\acs{Mc} \approx \SIrange{4.9E6}{2.2E8}{\Msun}$ for typical BCGs
\end{itemize} 
The smaller the dark halos (from dSphs to typical ellipticals to typical BCGs), the lesser their masses and the lesser their core compactness, and viceversa. This tendency ends at the larger (i.e. more extended) \ac{DM} halos, having a core of critical mass which is described in more detail by the continuous-magenta solutions (e.g. typical spirals, ellipticals and BCGs) in \cref{fig:profiles}. Additionally, the trend can be checked by comparing the group of values in columns $8$ with $5$ and $6$ among the different galactic structures in \cref{tbl:profiles}. 

The quantum core masses \acs{Mc}, the total halo masses \acs{Mtot} and the (consequent) associated window for the plateau densities \acs{rhop} (defined at the minimum of the \ac{RAR} rotation curve and inherent to each of the \ac{RAR} solutions), have to be considered as explicit predictions of the \ac{RAR} model.\footnote{
	The reader is referred to \cref{sec:analysis} for a full description of the limiting predicted properties of the RAR profiles in terms of the free model parameters.
} These predicted values are contrasted in more details within the context of the \acs{Mbh} - \acs{Mtot} relation and the constancy of the \textit{central} surface \ac{DM} density in \cref{sec:correlations}, as a consistency check of the model.

\subsection{Galaxy fitting examples: the case of Sculptor and UGC05986}

\subsubsection{The Sculptor dSph}
\begin{figure}%
	\centering%
	\includegraphics[width=\hsize]{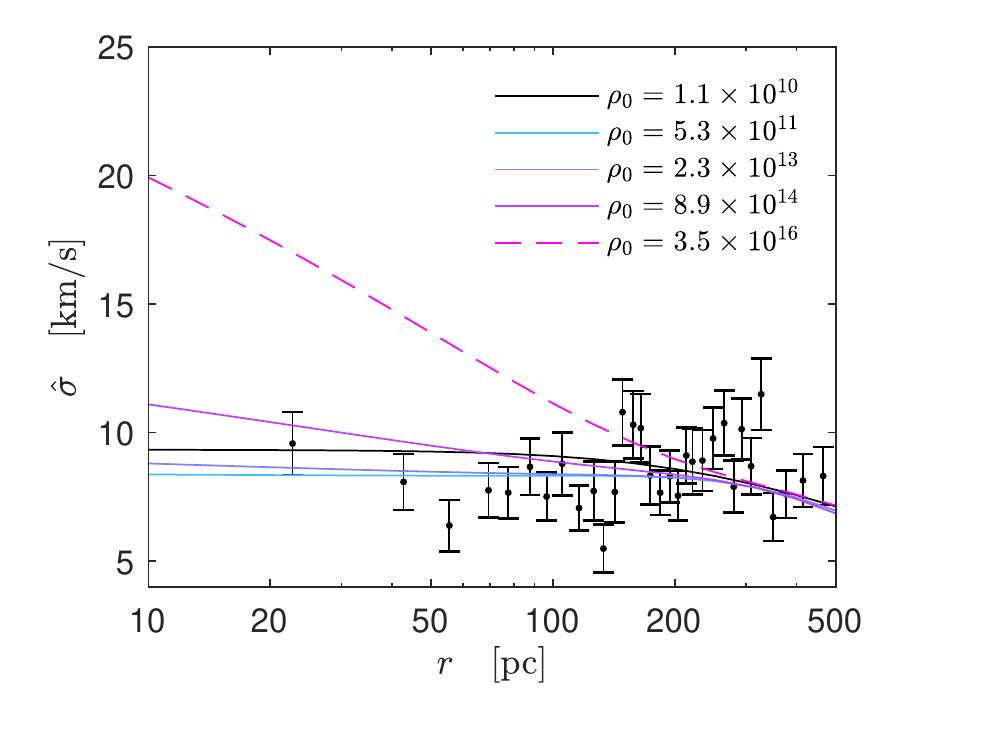}
	\caption{Comparison between the observed (projected) dispersion velocity (\acs{velDisp-proj}) of Sculptor, taken from \citep{2009ApJ...704.1274W}, against the same magnitude as predicted by a standard Jeans analysis. The latter uses the \ac{RAR} benchmark solutions for the mass distribution $\acs{M-dm}(r)$ as shown in \cref{fig:profiles}. Each solution is labeled with the central density in units of $\si{\Msun\parsec^{-3}}$. Notice the fast rise in the inner-halo region (at $\sim$ few $\SI{10}{\parsec}$) due to the cuspy \ac{RAR} halo with $\rho_0 = \SI{3.5E16}{\Msun\parsec^{-3}}$, implying a clear data mismatch in the case of this limiting dashed-magenta solution.}%
	\label{fig:sculptor-veldisp}%
\end{figure}

The aim is thus to link our typical RAR solutions for dSphs with a proper observable such as the (projected) dispersion velocity \acs{velDisp-proj} arising from Jeans analysis (as the one applied for dwarfs in \citealp{2009ApJ...704.1274W}) to be then compared with the corresponding data. We will consider for definiteness one of the best resolved MW satellites such as the Sculptor galaxy, as studied in \citealp{2009ApJ...704.1274W}. For this we assume our \ac{RAR} \ac{DM} mass profile $\acs{M-dm}(r)$ and a Plummer profile for the stellar (surface) density (with the corresponding $\acs{r-half}$ and orbital anisotropy for stellar components adopted in \citealp{2009ApJ...704.1274W}). This is done for the five different benchmark solutions ($mc^2= \SI{48}{\kilo\eV}$) as given in \cref{fig:profiles}. From this (back-of-the-envelope) comparison, which is shown in \cref{fig:sculptor-veldisp}, it turns out that while all of our solutions provide reasonable fits on halo scales\footnote{
	More refined fits to the data could be obtained within the \ac{RAR} model through a least squares analysis, when all halo observables are used to find the best fitting RAR free-parameters for $\acs{M-dm}(r)$, besides the two (generic) halo restrictions here applied (to appear in Paper III dedicated to dSphs).
} (somewhat similar to the cored-halo profile assumed in \citealp{2009ApJ...704.1274W}), some of them present a clear mismatch (of nearly a factor $2$) through the more central inner-halo scales (see dashed-magenta curve in \cref{fig:sculptor-veldisp}). This clear difference occurs for the solutions with central temperatures close to \acs{beta0-max}, i.e. the ones having exceedingly large Keplerian velocity cores reaching inner halo scales of few $\sim \SI{10}{\parsec}$. 

It is important to further emphasize that such \acs{velDisp-proj} data mismatch through the center, is only present for the RAR solutions developing a cuspy trend in $\rho(r)$, which in turn occurs only for the larger $\beta_0$ solutions. Interestingly, such trend arises because of the much small extension of dSphs respect to the larger galaxy types together with the astrophysical condition (presence of a maximum in the halo rotation curve) imposed to the solutions. Indeed, when such condition is no longer fulfilled (see saddle-point behaviour in $v(r)$ for dSphs in dashed-magenta in \cref{fig:profiles}), an unphysical fast rising trend in \acs{velDisp-proj} arises, evidencing the relevance of asking for such astrophysical condition in the rotation curve.


\def\ROOTPATH{figures/dwarfs-sculptor}


\subsubsection{The UGC05986 spiral}

In order to provide a complementary and detailed fit for a different galaxy type to the one shown in the above section, we have selected a DM dominated and well resolved spiral galaxy (UGC05986) from the Spitzer Photometry and Accurate Rotation Curves (SPARC) date base \citep{2016AJ....152..157L}. Specific information about each galaxy (i.e Hubble type, inclination etc) are provided in different files at \url{http://astroweb.cwru.edu/SPARC/}. We take for this galaxy the corresponding galactocentric radius $r$ and rotation curves $\SYMvel$, from the file \href{http://astroweb.cwru.edu/SPARC/Table2.mrt}{Table2.mrt}, as well as the baryonic contribution $\SYMvbar$, composed of a bulge ($\SYMvbulge$), disk ($\SYMvdisk$) and gas component ($\SYMvgas$). The bulge and disk components are inferred from surface brightness observations for a given mass-to-light ratio. In sum, the baryonic component is given by \begin{equation}
	\label{eqn:baryonic-sum}
	\SYMvbar^2 = \Upsilon_\mathrm{b}^{\phantom{2}} \SYMvbulge^2 + \Upsilon_\mathrm{d}^{\phantom{2}} \SYMvdisk^2 + \SYMvgas^2
\end{equation} With the corresponding mass-to-light \textit{ratio factors} $\Upsilon_b$ and $\Upsilon_d$ for bulge and disk respectively. Then the rotation curve for each component traces immediately its centripetal acceleration $\SYMacc = \SYMvel^2/r$.
We thus fit the inferred DM rotation curve, $\SYMvdark^2 = \SYMvobs^2 - \SYMvobs^2$, with the Levenberg–Marquardt (LM) algorithm to find a $\chi^2$ minima. The quantity $\chi^2$ is calculated by \begin{equation}
	\chi^2(\vec p) = \sum \limits_{i=1}^N \left[\frac{V_i - v(r_i,\vec p)}{\sigma_i}\right]^2
\end{equation} with $N$ the number of data points, $V_i$ is the set of circular velocity data, $r_i$ is the corresponding set of radius data, $v(r_i,\vec p)$ is the predicted circular velocity at radius $r_i$ for the model parameter vector $\vec p$ and $\sigma_i$ is the uncertainty for $V_i$. For the RAR model under consideration, $\vec p = (\theta_0, W_0, \beta_0, m)$, we vary the three free parameter ($\theta_0, W_0, \beta_0$) for a fixed particle mass $mc^2= \SI{50}{\kilo\eV}$ until the least square condition is satisfied, giving: $\beta_0=1.6\times 10^{-7}$, $\theta_0=36.0$ and $W_0=64.5$.

\begin{figure}%
	\centering%
	\includegraphics[width=\hsize]{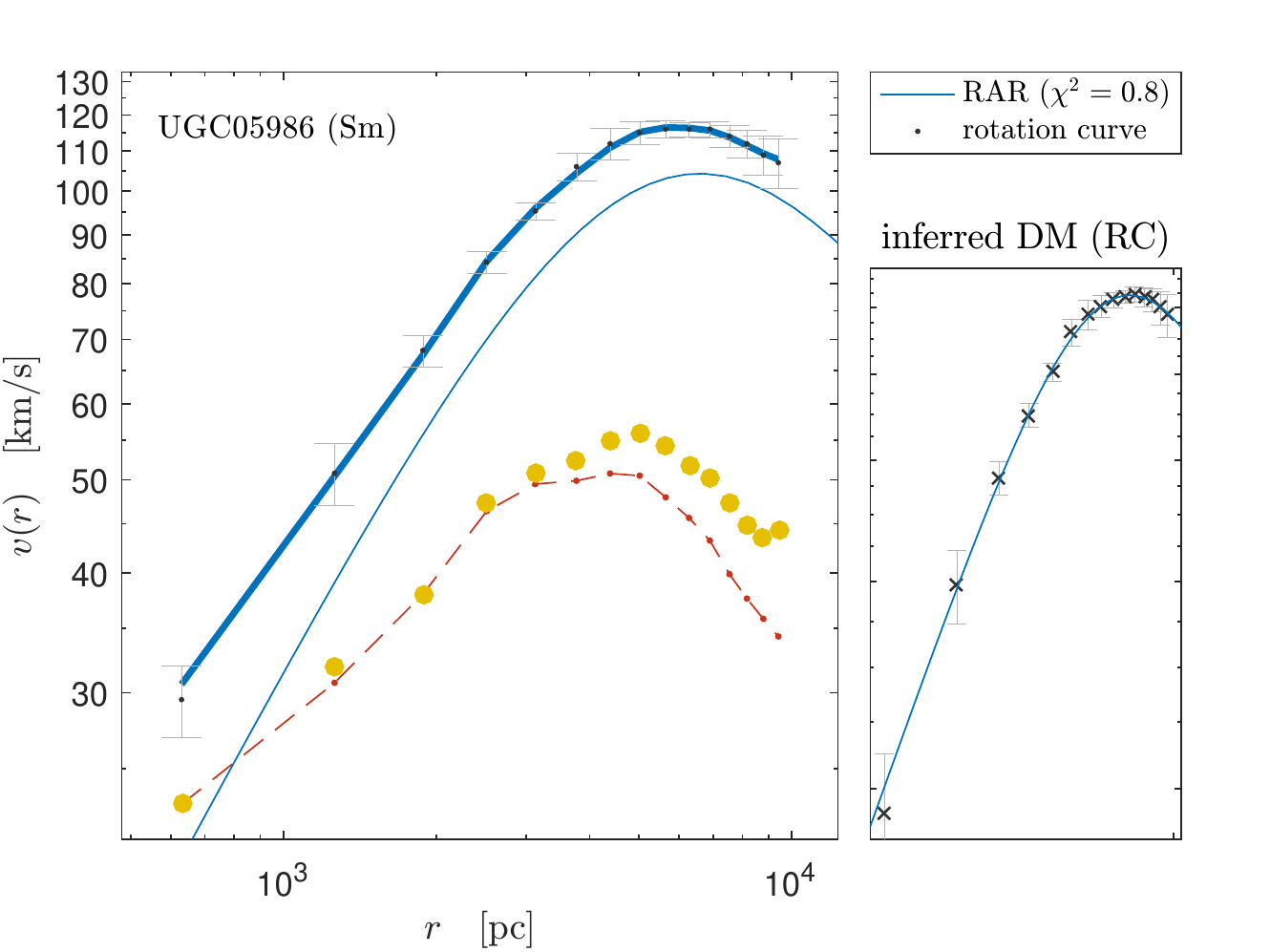}
	\caption{Best-fit of the RAR model for UGC05986, a dark matter dominated galaxy from the SPARC data base. Note also the well reproduced maxima of the total rotation curve (gray dots with error bars) as well as of the \ac{DM} component (solid blue). The baryonic (stars + gas) component is shown as big yellow dots while stars only are shown as red dashed line.}%
	\label{fig:SPARC:sample}%
\end{figure}

In \cref{fig:SPARC:sample} we show the rotation curves for different components of the galaxy UGC05986 as an example illustrating the necessity of cutoff effects for the \ac{DM} dominated halo\footnote{Notice moreover from \cref{fig:betathetaWZero} (typical spirals) that $W_0=64.5$ lays within the range of benchmark solutions which correspond with the sharp decreasing halo tails (i.e. $\rho\propto r^{-n}$ with $n>2$) in \cref{fig:profiles}}. A detailed analysis of the full SPARC galaxy sample will be presented in an accompanying paper.

\def\ROOTPATH{figures/sparcSample}




\subsection{Particle mass dependence}

In the case of typical dark halos in spiral, elliptical and BCGs a particle mass of $mc^2 = \SI{48}{\kilo\eV}$ provides the maximum (critical) core mass of $\acs{Mc-crit}=\ValMcoreCR$. If the mass is instead shifted to larger values, say $mc^2 \sim \SI{100}{\kilo\eV}$, a different three-parametric ($\theta_0,\beta_0,W_0$) family of solutions arises, able to reproduce the same \ac{DM} halo observables ($r_h, M_h$) for each case. But now the maximum (critical) core mass decreases to $\acs{Mc-crit}\sim \SI{E7}{\Msun}$ for galactic structures larger than dwarfs. These new solutions have exactly the same core-halo behavior as those in \cref{fig:profiles} with similar windows of \textit{predicted} core and total halo mass (\acs{Mc} and \acs{Mtot}) but ending at the lower critical core mass as indicated above.

More generally, the fermion particle mass range $mc^2 \approx \SIrange{48}{345}{\kilo\eV}$, as obtained from the Milky Way analysis in Paper I, implies stable \ac{DM} quantum cores with masses up to the critical values. The lower particle mass the higher the critical core mass. Thus, the corresponding range is $\acs{Mc-crit} \approx \SIrange{4.2E6}{2.2E8}{\Msun}$ due to the particle mass dependence $\acs{Mc-crit}\propto m^{-2}$ (see \citealp{2014IJMPD..2342020A}).

\subsection{About the critical core mass}

\begin{figure}%
	\centering%
	\includegraphics[width=\hsize]{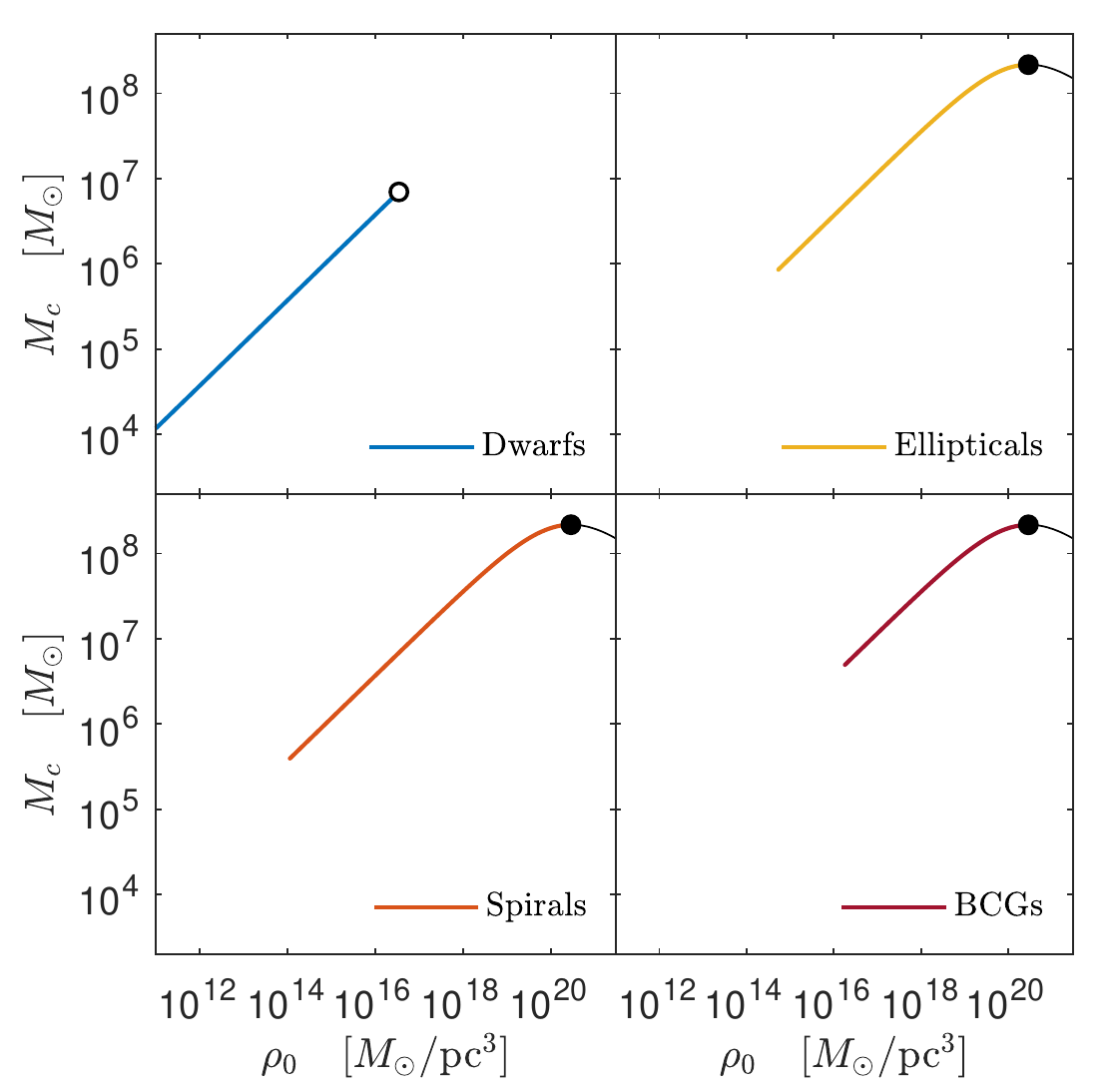}
	\caption{Explicit calculation of the $\rho_0$ - \acs{Mc} parameter space within the \ac{RAR} model for $mc^2 = \SI{48}{\kilo\eV}$ in the case of typical dwarf (blue line), spiral (red line), elliptical galaxies (yellow line) and BCGs (tamarillo line). Each type is described by given halo observable conditions (\acs{Mc} and \acs{rh}). Notice the case of spirals, ellipticals and BCGs where a critical DM core mass $\acs{Mc-crit}=\ValMcoreCR$ is reached at the maximum of the curve, when studied through the turning point criterion for core collapse in the $\rho_0$ - \acs{Mc} plane \citep[see e.g.][]{2014IJMPD..2342020A}.}%
	\label{fig:criticalcore}%
\end{figure}

The core of our configurations is supported by fermion degeneracy pressure. Therefore, the core is subjected to the gravitational instability leading to the concept of critical mass, traditionally introduced for white dwarfs \citep{1929ZPhy...56..851A,1930LEDPM...9..944S,1931ApJ....74...81C,1931MNRAS..91..456C,1932PhyZS...1..285L} and neutron stars \citep{1939PhRv...55..374O,1974PhRvL..32..324R}. When the pressure is dominated by the degeneracy pressure (i.e. when the temperature is much lower than the Fermi temperature), the fermion kinetic energy depends only on density, and so the equilibrium value of mass and radius of the configuration is set by the balance between the gravitational and the kinetic energy, at given finite density \citep{1939PhRv...55..374O}. This effect can already be seen in a post-Newtonian approximation determination of the equilibrium configurations  (see e.g.~\citealp{1974ApJ...189L..75W,1981A&A....97L..12C}). It can be shown that such a limiting configuration is the first turning-point of the equilibrium sequence of increasing central density, namely configurations along the sequence branch with $\tdiff{M}{\acs{rhoc}} > 0$ are stable, the ones with $\tdiff{M}{\acs{rhoc}} < 0$ are unstable, and $\tdiff{M}{\acs{rhoc}} = 0$ is the turning point, the critical mass configuration \citep[see][for a detailed discussion on the subject]{1983bhwd.book.....S}.
The above concept of the critical core mass \acs{Mc-crit} can be therefore formally achieved by finding the maximum (turning point) in a $\rho_0$ vs. \acs{Mc} diagram, as was shown in the context of the original \ac{RAR} model, i.e. for $W_0 \to \infty$ \citep[see][and references therein]{2014IJMPD..2342020A}. Labeled here as the \textit{critical-core} condition, this concept applies in the same way for the actual \ac{RAR} model with cutoff ($W_0 < \infty$), see \cref{fig:criticalcore}.

Accordingly, typical spiral and elliptical galaxies as well as BCGs reach the critical core mass (e.g the turning point at a critical density), corresponding to a critical temperature parameter $\acs{beta0-crit}$. For typical dwarfs, on the other hand, the maximal temperature ($\acs{beta0-max}< \acs{beta0-crit}$) is set by the astrophysical condition, such as requiring a maximum in the rotation curve on halo scales The (i.e. for $\beta_0 > \acs{beta0-max}$ such condition is broken as explicited in the dashed solutions of \cref{fig:profiles} for dwarfs). This limits the maximal core mass (and central density), being far away from the critical value.

\def\ROOTPATH{figures/analysis-core}

\subsection{Role of the degeneracy and cutoff parameters in the core-halo morphology of \ac{RAR} solutions}

The fact that higher compactness of the core is obtained by increasing the temperature parameter, while maintaining a rather low degeneracy, is understood by the semi-degenerate nature of our fermionic solutions in contrast with a fully degenerate regime. The combination of the finite-temperature effects and the monotonically decreasing change (from positive to negative values) of the degeneracy parameter with the distance (see e.g. upper panel of fig.~2 in \citealp{2018PDU....21...82A}) are the responsible for the clear \textit{dense quantum core - Boltzmannian tail} behavior of the density profiles in \cref{fig:profiles}. Once in the diluted regime, and for solutions which are still away from becoming critical, a small increase in $W_0$ has important effects on the \ac{RAR} halo scales: the larger the cutoff parameter $W_0$, the more massive and more extended the galaxy gets as can be directly checked in \cref{fig:profiles} and \cref{fig:betathetaWZero} from dSphs to ellipticals to BCGs, respectively.

This fact is better understood through the role of the escape energy $\varepsilon(r)$ in the distribution function (see {sec:RAR-model}). The larger the escape energy $\varepsilon(r) \neq \infty$, the larger the momentum (and energy) space the particles can occupy at any radius. Note, the escape energy is proportional to the cutoff parameter $W(r)$. In consequence, the solution cover more extended total spatial extensions before $W(r)$ reaches 0 at the boundary radius \acs{rb}.
\section{Galaxy parameter correlations}
\label{sec:correlations}

In the previous sections we have successfully compared and contrasted the solutions of the \ac{RAR} model with a wide range of galactic observables. We turn now to analyze if the \ac{RAR} model agrees with the following observational correlations:
\begin{itemize}
	\item The constancy of the \textit{central} surface \ac{DM} density in galaxies, e.g. $\acs{rho0D} \acs{r0} \approx 140_{-50}^{+80}\,\si{\Msun\parsec^{-2}}$. It spans about 14 orders of (absolute) magnitude (\acs{MB}), where \acs{rho0D} and \acs{r0} are the \textit{central} \ac{DM} halo density at the one-halo-scale-length of the Burkert profile \citep{2009MNRAS.397.1169D}.
	\item The \acs{Mbh} - \acs{Mtot} relations with \acs{Mbh} the mass of the compact dark object at the centre of galaxies and \acs{Mtot} the total \ac{DM} halo mass \citep{2002ApJ...578...90F,2011Natur.469..377K,2015ApJ...800..124B}.
\end{itemize}

In order to show this, we use the full family of astrophysical \ac{RAR} solutions (i.e. contained within the blue-shaded region of \cref{fig:profiles}) for typical dSphs, spiral and elliptical galaxies as well as typical BCGs. Such solutions cover the maximal free parameter space ($\beta_0$,$\theta_0$,$W_0$) for each galaxy type as constrained by the halo observables $(\acs{rh},\acs{Mh})$ for the particle mass $mc^2 = \SI{48}{\kilo\eV}$ together with the \textit{quantum core} condition. Correspondingly, a well defined window of predicted masses ($\acs{Mc},\acs{Mtot}$) is obtained (see \cref{fig:ParameterAnalysis} for details). As we show below, the knowledge of the corresponding values of the plateau density \acs{rhop} is also important for the analysis of the \textit{central} surface \ac{DM} density relation.

We can proceed now to make a consistency check of the predictions of the \ac{RAR} model by contrasting them within the physical observed spread of the correlations. Notice that the constancy of the \textit{central} surface \ac{DM} density deals only with \ac{DM} halos while the \acs{Mbh} - \acs{Mtot} relations correlate both, the central and total halo dark object masses. Traditionally, the central compact dark objects are assumed as \acp{SMBH}. But here we interpreted them as \ac{DM} quantum cores with the exception of active galaxies harboring supermassive central objects above $\sim\SI{2E8}{\Msun}$.

\subsection{\ac{DM} surface density relation}
\label{sec:correlations:donato}

Regarding the \textit{central} surface \ac{DM} relation, we first take from the literature the values for the blue absolute magnitude \acs{MB}, corresponding to each typical galaxy within each galaxy type considered above. Thus we adopt 
$\acs{MB}\approx -10.2$ for typical dSphs \citep{1995MNRAS.277.1354I},
$\acs{MB}\approx -20.8$ for the Milky Way \citep{2004AJ....127.2031K},
$\acs{MB}\approx -20.5$ for typical spirals \citep{2008AJ....136.2648D},
$\acs{MB}\approx -21.5$ for typical ellipticals \citep{2005ApJ...635...73H} and
$\acs{MB}\approx -23$ for typical BCGs \citep{2013ApJ...767..102W}.

Then, in order to calculate the \ac{DM} surface density $\acs{Sigma0D}=\acs{rho0D} \acs{r0}$ in each case, we simply realize that the equivalent of the Burkert central density \acs{rho0D} would correspond to the density of the plateau \acs{rhop} within the \ac{RAR} model. The relation between both one-halo scale lengths is given by $\acs{r0}\approx 2/3 \acs{rh}$, where \acs{rh} is fixed for each galaxy type according to the imposed halo constraints. For the corresponding family of \acs{rhop} values see \cref{tbl:profiles} and \cref{fig:ParameterAnalysis}.

We thus calculate the product $2/3 \acs{rhop} \acs{rh}$ for each theoretical profile in the case $mc^2=\SI{48}{\kilo\eV}$, including the Milky Way, and finally contrast the pair ($\acs{MB},\acs{Sigma0D}$) with the observational relation found by \citet{2009MNRAS.397.1169D}.

\begin{figure}%
	\centering%
	\includegraphics[width=\hsize]{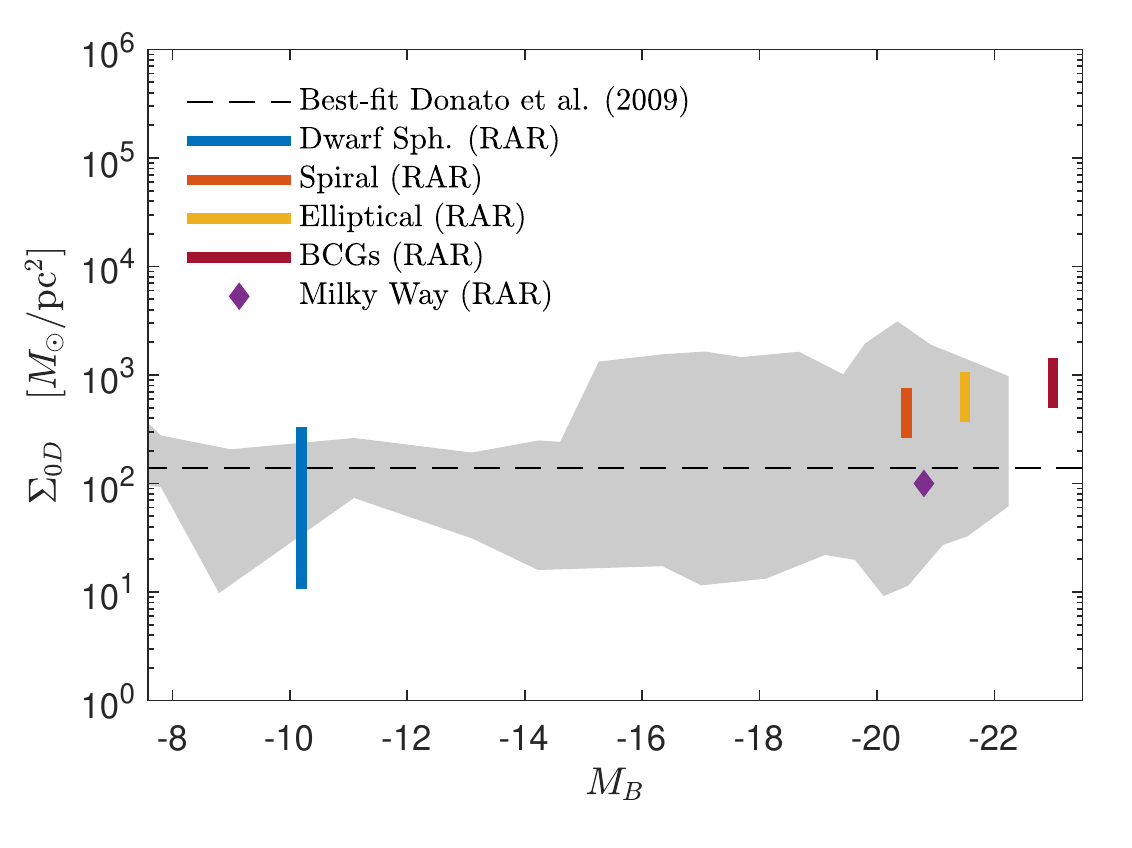}
	\caption{The surface \ac{DM} density as predicted by the \ac{RAR} model (see vertical colour lines) for each galactic structure in correspondence with the astrophysical solutions (i.e. blue-shaded regions in \cref{fig:profiles}). The dashed horizontal line represents the Universal relation from the best fit of the data as found by \cite{2009MNRAS.397.1169D}. The dark-gray region indicates the delimited area by the $3-\sigma$ error bars of all the data points. The result shows the ability of the three parametric \ac{RAR} model (for $mc^2 = \SI{48}{\kilo\eV}$) to be in agreement with the DM surface density observations (see text for further details regarding BCGs).}%
	\label{fig:donato}%
\end{figure}

The results for dwarfs to ellipticals are in very good agreement with the observed relation, see \cref{fig:donato}. For simplicity, the latter is displayed as the overall dark-grey region delimited (or enveloped) within the $3-\sigma$ error bars along all the data points considered in \citet{2009MNRAS.397.1169D}. The predicted surface density (vertical solid lines), for the adopted \acs{MB} values, are located within the expected $3-\sigma$ data region. The magnitude \acs{MB} of typical BCGs is beyond the observed window reported in \citet{2009MNRAS.397.1169D} who considered up to elliptical structures, but their predicted values are somewhat similar to the latter.

Remarkably, our results of the \ac{RAR} model show a mild increasing behavior with decreasing \acs{MB}. This trend resembles the analogous universal relation presented in \citet{2009arXiv0911.1774B} where larger elliptical galaxies as well as clusters were included in the analysis, contrary to the sample presented in \citet{2009MNRAS.397.1169D}.

\def\ROOTPATH{figures/donato}

\subsection{Super massive central object - DM halo mass relation}

\def\ROOTPATH{figures/ferrareseCompact}\begin{figure*}[t!]%
	\centering%
	\includegraphics[width=0.5\hsize]{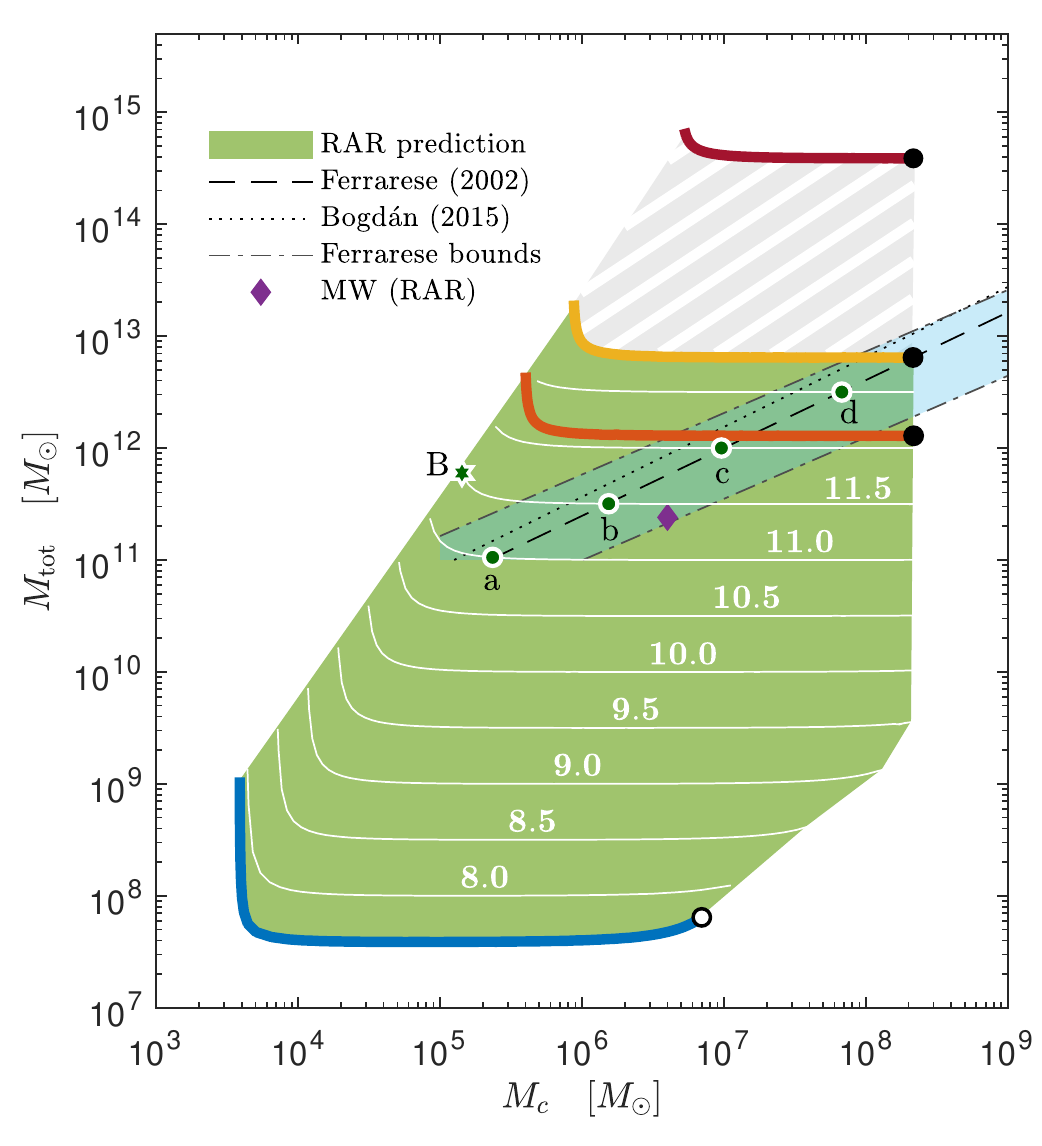}%
	\includegraphics[width=0.5\hsize]{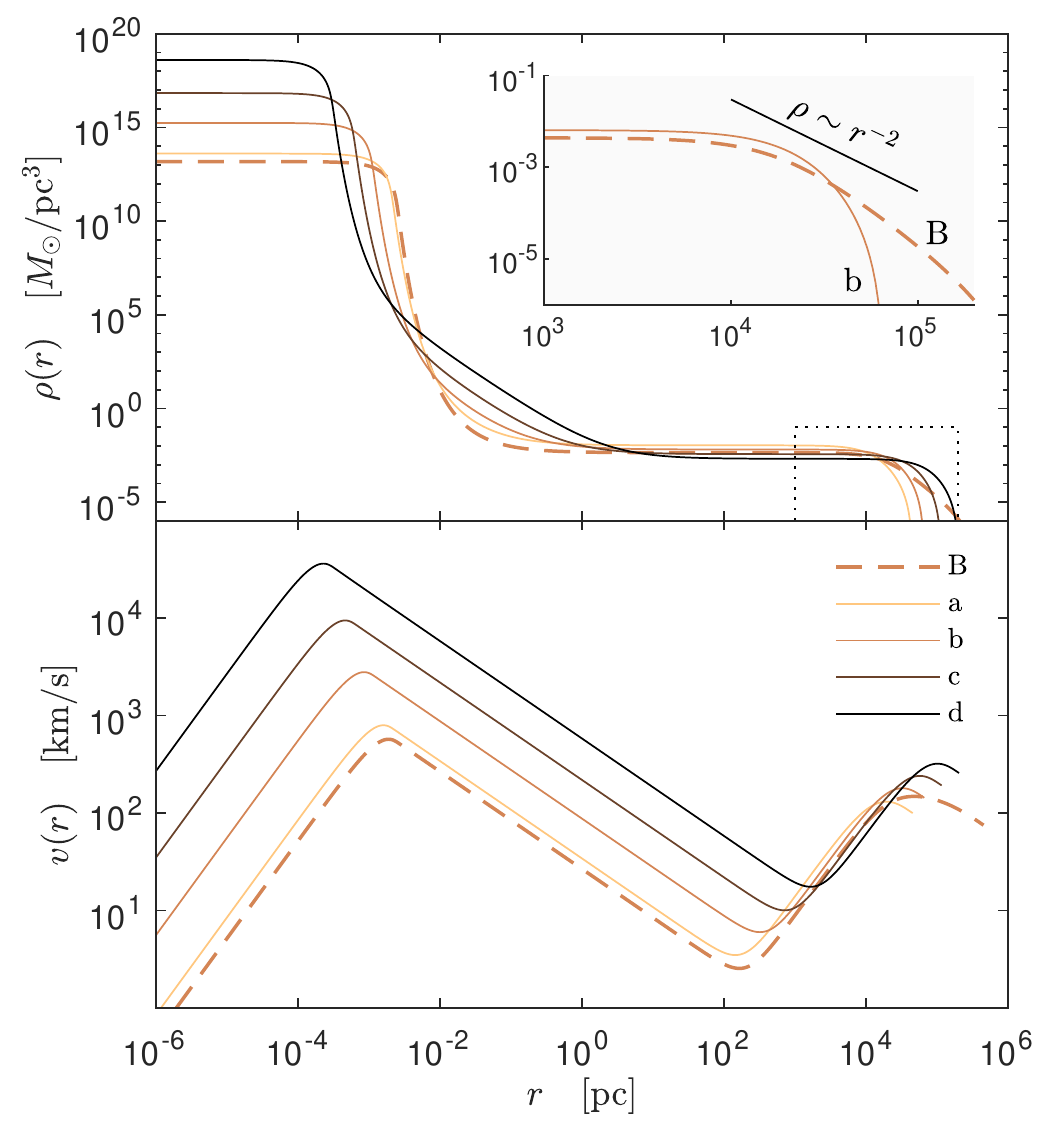}%
	\caption{(\textbf{left}) Prediction of the \acs{Mc} - \acs{Mtot} relations within three parametric \ac{RAR} model (for $mc^2 = \SI{48}{\kilo\eV}$). The different coloured lines read for each galaxy type in correspondence with the astrophysical \ac{RAR} solutions as given in \cref{fig:profiles}. The green area, on the other hand, covers all RAR predictions for a given halo mass in the range $\acs{Mh} \approx \SIrange{E7}{E12}{\Msun}$ and fulfilling $\frac23\acs{rhop}\acs{rh} = \SI{140}{\Msun/\parsec^2}$ as inferred from the Donato relation. The white lines show a set of families with given halo mass \acs{Mh} (and labeled by the \acs{Mtot} value in the horizontal regime). The results show the ability of the RAR model to be in agreement with the different \acs{Mbh} - \acs{Mtot} relations, as considered in the literature, and explicited in the blue-ish stripe. The shaded area above ellipticals is just the extrapolation of the \acs{Mc} - \acs{Mtot} RAR prediction for BCGs obtained assuming the constant (best fit) of the Donato relation (see also \cref{sec:rhMh-relation}). The filled-black dots correspond to the critical core mass \acs{Mc-crit}, and the empty-black dot indicates the limiting maximum core mass \acs{Mc-max} for dwarfs. %
	(\textbf{right}) Density and velocity profiles associated with the 4 benchmark solutions of the left panel labeled by (a-d). Such solutions lie along the best-fit of \citet{2002ApJ...578...90F} with the given mean surface density $\frac23\acs{rhop}\acs{rh} = \SI{140}{\Msun/\parsec^2}$. While the dashed curve (B) representing an isothermal-like RAR solution (corresponding with dot B in left-panel) is clearly disfavored by the observed correlation, the continuous curves (as e.g. (b)) are favored.}%
	\label{fig:McMDM}%
\end{figure*}
Concerning the \acs{Mbh} - \acs{Mtot} relations, we show in left-panel of \cref{fig:McMDM} the predicted $(\acs{Mc},\acs{Mtot})$ values (see also \cref{fig:ParameterAnalysis}) for many different astrophysical family of RAR solutions including (but not limited to) the typical galactic structures of \cref{fig:profiles}), together with the observationally inferred best-fit relations found in \citet{2002ApJ...578...90F,2015ApJ...800..124B}. The observational inferred relations are limited to the region where data supports, i.e. the \textit{Ferrarese strip denoted by blue-ish area} in left-panel of \cref{fig:McMDM}. Interestingly such observational strip is explicitly shown to be contained (up to $\acs{Mc}\sim \SI{E8}{\Msun}$) within the larger \ac{RAR} green-ish predicted area from small spirals up to ellipticals. The red and yellow continuous thick L-shaped lines correspond to the typical spirals and ellipticals considered in \cref{sec:constraints}. While the thin white lines (labeled with different \acs{Mtot} values) cover many different one-halo scale observables (\acs{rh},\acs{Mh}) \textit{besides} the ones associated to the typical galactic structures introduced in \cref{sec:constraints}. Such new (\acs{rh},\acs{Mh}) values are obtained from the \textit{independent} Donato relation as explained in \cref{sec:rhMh-relation} and \cref{fig:MhRh}, with the aim to maximally cover the $(\acs{Mc},\acs{Mtot})$ plane. The RAR predicted (green-ish) area extends out of the Ferrarese strip, indicating a potential window of core-halo masses which are either not yet observed or unphysical. 

Indeed, RAR solutions deviating too far-left of the observed strip (e.g. dot B in left-panel of \cref{fig:McMDM}), has isothermal-like density tails ($\rho(r) \sim r^{-2}$) as plotted in right-panel, at difference with RAR solutions lying along the best-fit relation (e.g. dot b, see both panels). The later has halo tails decreasing as $\rho(r)\propto r^{-3}$ somewhat inside $r_h$ \footnote{
	Due to the cut-off effects, the RAR halo tails behaviour for solutions which are favored by the correlation are of polytropic nature, with $\rho \sim r^{-4}$ for $r=\acs{rh}$.
} more in line with other phenomenological profiles arising from N-body simulations either in WDM or CDM cosmologies. From this comparison we can conclude that the observed $(\acs{Mc},\acs{Mtot})$ correlation \textit{disfavors} RAR solutions showing isothermal-like halo tails, while favors more sharply decreasing halo tails. The Milky Way \ac{RAR} solution is also plotted for completeness, showing a good agreement as well.

The case of typical dwarf galaxies (thick blue curve) is located at the lower-left end of the $(\acs{Mc},\acs{Mtot})$ plane in \cref{fig:McMDM}, beyond the observationally inferred blue-ish strip. It is worth to stress that no observational data exist yet in that part of the correlation and thus the obtained results are considered as a pure RAR model prediction, subject to observational scrutiny. However, special attention has to be given to our dwarf galaxy predictions in view of the recent observational reports of putative massive black holes of few $\sim \SI{E6}{\Msun}$ detected in ultra-compact dwarf galaxies of total mass of few $\sim \SI{E7}{\Msun}$ \citep[e.g.][]{2014Natur.513..398S,2017ApJ...839...72A,2018ApJ...858..102A,2018MNRAS.477.4856A}. Similarly interesting candidates are the recent discovered \acp{IMBH} in globular clusters \citep[e.g.][]{2017Natur.542..203K,2017MNRAS.468.2114P}.

When we look at larger structures beyond ellipticals we find BCGs as further convenient predictions. Interestingly (and different to the other smaller structures considered) their dark central cores are limited by a relatively light-mass value ($\acs{Mc-crit} \sim \SI{2E8}{\Msun}$) when compared to the very massive halo of few $\SI{E14}{\Msun}$. Having halo properties very similar to typical ellipticals and spirals, in line with the universal DM halo morphology within the standard picture of hierarchical structure formation.

Notice that a vertical trend up to infinity (not shown) in the L-shaped lines of \cref{fig:McMDM} corresponds to \ac{RAR} solutions having the minimum core mass (or minimum $\rho_0$) which in turn implies the larger cutoff parameters until $W_0\rightarrow\infty$ with no particle escape whatsoever (see \cref{fig:ParameterAnalysis} for the case of typical galactic structures). Those solutions develop more extended and more massive halos. Such trend ends in an infinite halo mass when the limiting $\rho\propto r^{-2}$ isothermal \ac{RAR} tail is reached, unless some (extra) virial condition is imposed to them (e.g. typically setting a minimum threshold density at about $\sim 200 \acs{rho-crit}$ to any profile, where \acs{rho-crit} is the critical density of the Universe). Solutions with total masses exceeding such virial constraint are excluded as astrophysical ones and therefore not appearing in \cref{fig:McMDM}.

It is appropriate to stress at this point that the \acs{Mtot} values appearing in the observed relations plotted in \cref{fig:McMDM}, were calculated at the virial radius within a \acl{NFW} \ac{DM} model \citep{2002ApJ...578...90F}. While in our case, they were obtained at the surface radius \acs{rb} of the \ac{RAR} model equilibrium configurations. Besides, the majority of the values of \acs{Mbh} in the observed \acs{Mbh} - \acs{Mtot} relation have been obtained through the so-called $\acs{Mbh} - \acs{velDispB}$ relation, with \acs{velDispB} the bulge dispersion velocity \citep{2000ApJ...539L...9F,2000ApJ...539L..13G,2009ApJ...698..198G}. However, in the case of dwarf galaxies the observational inference of the dark central object mass via the dispersion velocity is unclear \citep[see, e.g.,][and references therein]{2005ApJ...628..137V,2011Natur.469..377K}. Interestingly, \citet{2005ApJ...628..137V} attempted to give an estimate of such a central object mass in the case of NGC205 obtaining $\acs{Mbh}\sim \SIrange{E3}{E4}{\Msun}$ (interpreted there as an \ac{IMBH}), which is in agreement with the values of \acs{Mc} for the typical dSphs analyzed in this work (see \cref{tbl:profiles}).

Finally, in the case of larger elliptical galaxies, it is interesting to note that the maximum quantum core mass $\acs{Mc-crit} \sim \SI{2E8}{\Msun}$ (for $mc^2=\SI{48}{\kilo\eV}$) predicted by our model, is in striking consistency with the uppermost (sample-representative) central mass \acs{Mbh} obtained in \citet{2015ApJ...800..124B}, from an X-ray imaging analysis of more than 3000 isolated and without AGN activity elliptical galaxies. These results, when viewed through our theoretical \acs{Mbh} - \acs{Mtot} relation, give support to our idea that normal elliptical galaxies may harbor dark central objects (not yet BHs) without showing AGN-like activity, while larger SMBHs masses, do show AGN properties, reaching the upper end of the \acs{Mbh} - \acs{Mtot} relation.


Additional verification of the above predictions of the \ac{RAR} model needs the observational filling of the gaps in the $(\acs{Mc},\acs{Mtot})$ plane from dwarfs all the way up to ellipticals. This has been partially done for disk galaxies from the SPARC data base and will be presented in an accompanying paper.

\section{Concluding remarks}
\label{sec:conclusion}

In Paper I we clearly demonstrated that gravitationally bounded systems based on fermionic phase-space distributions including for escape velocity effects and central degeneracy, can explain the \ac{DM} content in the Galaxy while providing a natural alternative for the central BH scenario in \acs{SgrA}. This highly compelling result is bolstered here by the analysis of different galactic structures ranging from dwarfs to ellipticals to galaxy clusters.

As an interesting example, we have discussed in this paper the solutions for $mc^2=\SI{48}{\kilo\eV}$, where the model is able to explain the \ac{DM} halos from typical dSph to normal elliptical galaxies up to typical BCGs, and predict the presence of massive compact dark objects from $\sim \SI{E3}{\Msun}$ up to $\sim \SI{E8}{\Msun}$ at their respective centers. A key point of the present \ac{RAR} model is the ability to fulfill the observed properties of galaxies, such as the \acs{Mbh} - \acs{Mtot} and the $\acs{Sigma0D} \approx {\rm const}$ universal relations, for a unique \ac{DM} fermionic mass. The versatility of the physical three-parametric solutions, can also account for the (real) physical spread observed in the correlation between dark halo mass Vs. dark central object mass, as observationally inferred in \citet{2002ApJ...578...90F,2011Natur.469..377K,2015ApJ...800..124B}. Whether or not such full window of compact dark-object masses at the centres of galaxies occur in Nature is a theme of future observational works, particularly interesting in the case of the smallest (i.e. faintest) dwarf galaxies.

Nevertheless, the analysis should cover all observed plethora of galactic dark halos with corresponding dark compact central objects. In particular, the RAR model predicts that galaxies with similar halo properties (i.e. total halo mass and radius) can harbor different dark quantum core masses, spanning up to about $3$ orders of magnitude. This peculiar feature is an important result of our theory, considering that very similar Seyfert-like galaxies have been observed to shown values of \acs{Mc} that can differ in nearly one order in magnitude ($\acs{Mc}\sim \SIrange{1E7}{8E7}{\Msun}$, see \citealp{2010ApJ...721...26G}). 

At the upper limit of the compact core mass range, it provides, on astrophysical basis, possible clues on the formation of \acp{SMBH} in galactic nuclei. In the case of typical dark halos in normal spiral and elliptical galaxies, self-gravitating fermions with a particle mass of $\SI{48}{\kilo\eV}$ may produce a maximum (critical) core mass of $\acs{Mc-crit}=\ValMcoreCR$, at the onset of gravitational collapse.

The majority of the supermassive dark central objects are comprised within $\acs{Mc}\sim \SI{E8}{\Msun}$ \citep{2009ApJ...698..198G}. However, in the largest elliptical galaxies have been also observed more massive dark objects up to $\sim \SI{E10}{\Msun}$. Those candidates are most likely \acp{SMBH} associated with active galaxies and are characterized by a clear X-ray and radio emissions as well as jets. Such \acp{SMBH} may be explained starting from a BH seed of mass \acs{Mc-crit} formed out of the collapse of our critical quantum \ac{DM} cores. After its formation, such a BH seed might start a baryonic and/or dark matter accretion process from their massive galactic environment ($M_g\sim \SI{E12}{\Msun}$). An accretion of $\sim \SI{1}{\percent}$ of the (inner) baryonic mass of the galaxy onto the \acs{Mc-crit} core mass obtained here, would be enough to explain the formation of the largest ($\MBH\sim \SIrange{E9}{E10}{\Msun}$) \ac{SMBH} masses.

Other observational data-sets such as Ly$\alpha$ forest constraints, or high resolution rotation curves of disk galaxies (points (b) and (c) mentioned in the introduction), are also important to discriminate between different particle-motivated \ac{DM} halo models such as the RAR model, \ac{FDM} or sub-keV degenerate fermions. Standard \ac{FDM} as the ones recently considered in \citet{2017PhRvD..95d3541H}, has an associated matter power spectrum which share the same features with the one of the \ac{WDM} paradigm at large scales, both exhibiting a clear drop (though somewhat different in shape) in power below a given scale~\citep{2014MNRAS.437.2652M}. Such similar behaviour allowed \citet{2017PhRvD..95d3541H} to roughly relate the lower bound mass values of \ac{FDM} to the ones of a \ac{WDM} fermion obtained by \citet{{2013PhRvD..88d3502V}} from Ly$\alpha$ forest data. In particular, a lower limit of $\SI{2.5}{\kilo\eVcc}$ (at $3-\sigma$) of the hypothetical \ac{WDM} particle (thermal relics) translates into $m_bc^2 > \SI{E-21}{\eV}$ for the ultra light boson at the same confidence level \citep{2017PhRvD..95d3541H}. Unfortunately, this bound strongly disfavours the typical \ac{FDM} range of masses needed to solve different small-scale issues of the CDM paradigm \citep{2017PhRvD..95d3541H}. Improved analysis making use of dedicated cosmological simulations within \ac{FDM} paradigm to account for the physics at the scales probed by the Ly$\alpha$ forest data (from the BOSS survey), exclude the mass range $m_b = \SIrange{E-22}{2.3E-21}{\eV}$ \citep{2017MNRAS.471.4606A}. 

In the case of fermionic models, it is then clear from the above paragraph that the $\sim$ few \si{\kilo\eV} lower bound found in \citet[][see also~\citet{2017JCAP...06..047Y} for more general lower bounds above \si{\kilo\eV} including for sterile neutrino \ac{WDM}]{2013PhRvD..88d3502V} is in tension with the below-\si{\kilo\eV} Thomas-Fermi DM models presented in \citet{2013NewA...22...39D,2015JCAP...01..002D,2017MNRAS.467.1515R}, while it remains in agreement with the \ac{RAR} model here presented.

Reproducing the (total) Milky Way rotation curve features posses as well serious challenges to \ac{FDM} models in the desired mass range needed to account for other astrophysical and small-scale cosmological observables \citep{2018arXiv180500122B}. Indeed, it was there explicitly shown that the solitonic core arising typically at bulge-scales in the case of typical disk galaxies, practically excludes the \ac{FDM} mass range $m_b = \SIrange{E-22}{E-21}{\eV}$. Finally, the case for highly degenerate fermionic halos (i.e. Thomas-Fermi models DM cited above), are not applicable at all to reproduce the rotation curves of disk galaxies (or larger), since they were originally motivated to fit the dispersion velocities of small dSphs \citep{2013NewA...22...39D} due to the predicted highly compact and small sized halos. Moreover, a closer look on such models (as the one recently analyzed in~\citet{2015JCAP...01..002D}) shows that they are disfavoured even to fit dSph dispersion velocities, and the artificial introduction of an isothermal tail is needed \citep{2017MNRAS.467.1515R}.

Thus, the results described in the above paragraphs put the \ac{RAR} model with fermion masses above $\SI{10}{\kilo\eV}$, in a position of privilege with respect to other particle-based models such as \ac{FDM} or the ones based on sub-keV fully-degenerate fermions. 

Finally, and in order to attempt to answer the question why nature constrains the \ac{RAR} model free parameters to the specific values shown in \cref{fig:betathetaWZero} and \cref{tbl:profiles}, it would need the extension of the present analysis into a broader theoretical context such as the formation and evolution of galaxies.

Such an insight may be likely gained through a detailed study of a dynamical theory of collisionless relaxation, including a full statistical/thermodynamical analysis of the condition under which (quasi) \ac{DM} halo relaxation is reached. Considerations based on maximization entropy approaches for given total mass (with corresponding running of the free model parameters), as the one analyzed in similar self-gravitating systems in \citet{2015PhRvD..92l3527C} (and references therein), could help in this direction. These interesting aspects are out of the scope of the present work and are subject of future research.
\section*{Acknowledgments}

We thank the referee for her/his very constructive and clear suggestions.
C.R.A acknowledges support by the International centre for Relativistic Astrophysics Network (ICRANet) and CONICET-Argentina.
J.A.R acknowledges support from the International Cooperation Program CAPES-ICRANet financed by CAPES-Brazilian Federal Agency for Support and Evaluation of Graduate Education within the Ministry of Education of Brazil.
A.K. is supported by the Erasmus Mundus Joint Doctorate Program by Grants Number 2014--0707 from the agency EACEA of the European Commission.

\appendix
\section{Parameter space analysis}
\label{sec:analysis}
Here we show how the halo observable constraints (\acs{rh} and \acs{Mh}) together with the additional \textit{quantum-core} condition define the limiting values of the free sets of \ac{RAR} configuration parameters ($\beta_0$,$\theta_0$,$W_0$) for the different galaxy types. Specifically \cref{fig:betathetaWZero} shows the full curves in the ($\beta_0, \theta_0, W_0$)-space for $mc^2=\SI{48}{\kilo\eV}$. Each galaxy is represented through a coloured $1$-dimensional line, i.e. in thick blue (dwarfs), red (spirals) and yellow (ellipticals), while typical BCGs are represented through a thick tamarillo line. We also include along each line the sets of the $5$ benchmark \ac{RAR} solutions, given in \cref{fig:profiles}, through dots in corresponding colours. This correspondence shows clearly the ranges of $\beta_0$, $\theta_0$, and $W_0$, encompassing all the astrophysical \ac{RAR} solutions.

\subsection{About the 1-dimensional curves of free \ac{RAR} model parameter space}
The fact that the halo scale radius \acs{rh} sets a specific morphological point in the \ac{RAR} solutions (i.e. as in $\acs{v-dm}(r)$, $\acs{M-dm}(r)$, $\beta(r)$, $\theta(r)$, etc.), it must necessarily depend on the specific choice of the initial conditions, i.e. $$\acs{rh} \equiv \acs{rh}(\beta_0, \theta_0, W_0, m)$$ This functional dependence, together with $$\acs{Mh} \equiv \acs{M-dm}(\beta_0, \theta_0, W_0, m, \acs{rh})$$ clearly defines a $1$-dimensional curve in the ($\beta_0, \theta_0, W_0$) \ac{RAR} configuration parameter space once $m$, \acs{Mh} and \acs{rh} are given (i.e. $4$ free parameters and $3$ constraints).

The number of free parameters of the model may be reduced to three when the particle mass $m$ is set (i.e. in the range $mc^2 \approx \SIrange{48}{345}{\kilo\eV}$ as obtained in PaperI). This approach requires only $2$ constraints, such as \acs{rh} and \acs{Mh}. If instead only one constraint (i.e. $M(r)$ with $r\neq \acs{rh}$ a hypothetically well constrained halo scale-length) is applied instead of the two constraints used in this work, then a narrow $2$-dimensional region would arise in the ($\beta_0, \theta_0, W_0$)-space of \cref{fig:betathetaWZero}. Nevertheless, many of the solutions in this (more) extended family will certainly provide worst fits to the baryonic data (as e.g. for the case of \acs{velDisp-proj} in dSphs) than the solutions here presented, considering less observable constraints were used.

\def\ROOTPATH{figures/analysis-main}\begin{figure*}%
	\centering%
	\includegraphics[width=\hsize]{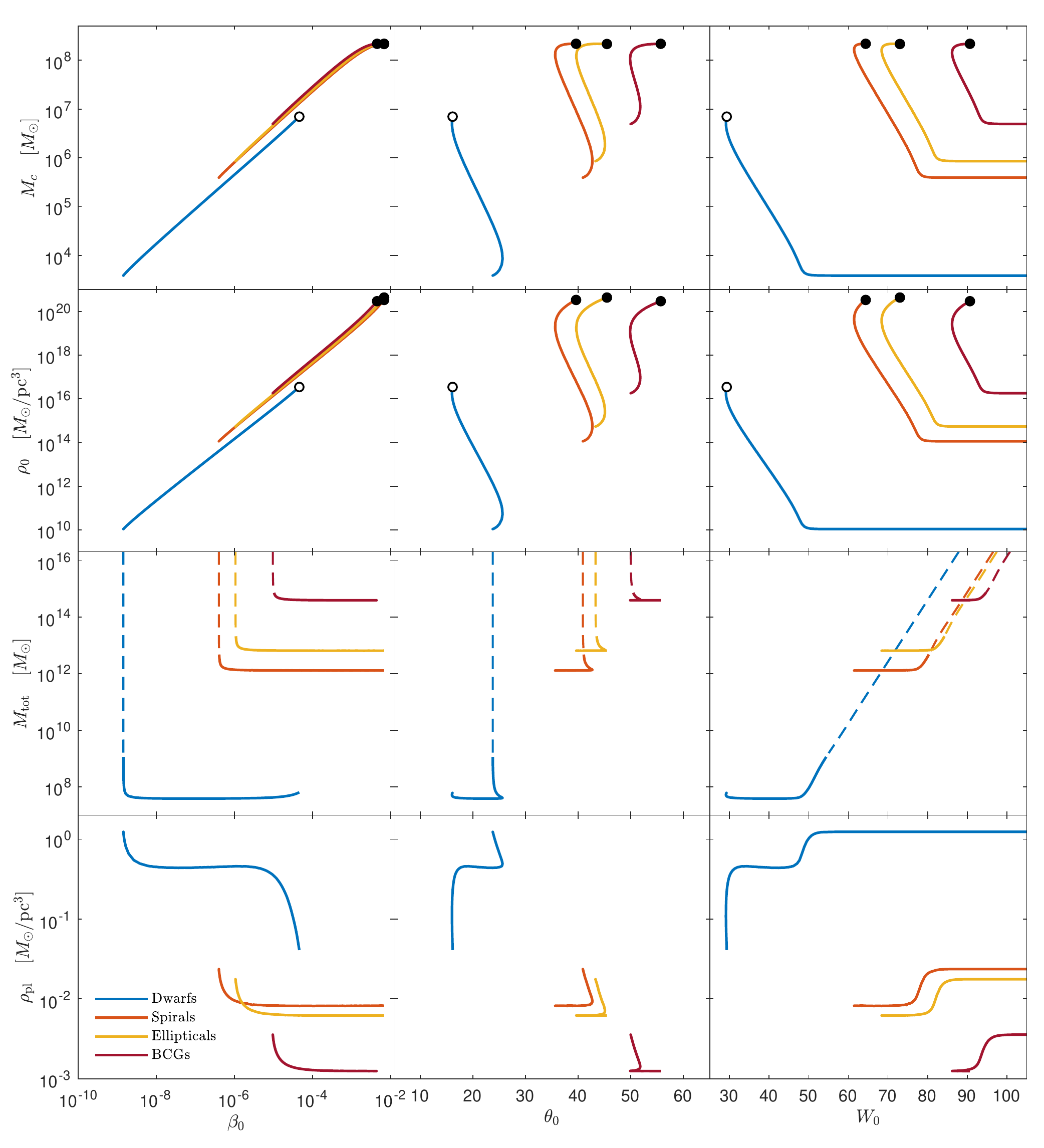}
	\caption{Full display of \ac{RAR} magnitudes ($\rho_0$,\acs{Mc},\acs{Mtot}, and \acs{rhop}) for $m c^2=\SI{48}{\kilo\eV}$ as a function of $\beta_0$ (left column), $\theta_0$ (central column) and $W_0$ (right column). The dashed-lines in the \acs{Mtot} row correspond with isothermal-like halo tails having densities \textit{below} the virial value $\sim 200 \acs{rho-crit}$ as set in \cref{sec:correlations}. Shown are three typical galaxy types (dwarfs, spirals, ellipticals) and typical BCGs representing larger structures. Each case is constrained by halo observables (\acs{rh} and \acs{Mh}). The existence of a critical core mass of $\acs{Mc-crit}=\ValMcoreCR$, for the case of typical spiral and elliptical galaxies as well as BCGs, is denoted by a filled black dot. In the case of dwarfs there is only a maximum value \acs{beta0-max} (with associated \acs{Mc-max}) and denoted by an empty black dot. Notice also the larger RAR plateau density ($\sim \SI{E-1}{\Msun/\parsec^3}$) for the dwarf galaxies (on inner-halo scales) with respect to the corresponding lower values for larger galactic structures (in line with observations, see e.g. \citet{2009ApJ...704.1274W}).}%
	\label{fig:ParameterAnalysis}%
\end{figure*}
\def\ROOTPATH{figures/analysis-parameter}\begin{figure*}[t!]%
	\centering%
	\includegraphics[width=0.7\hsize]{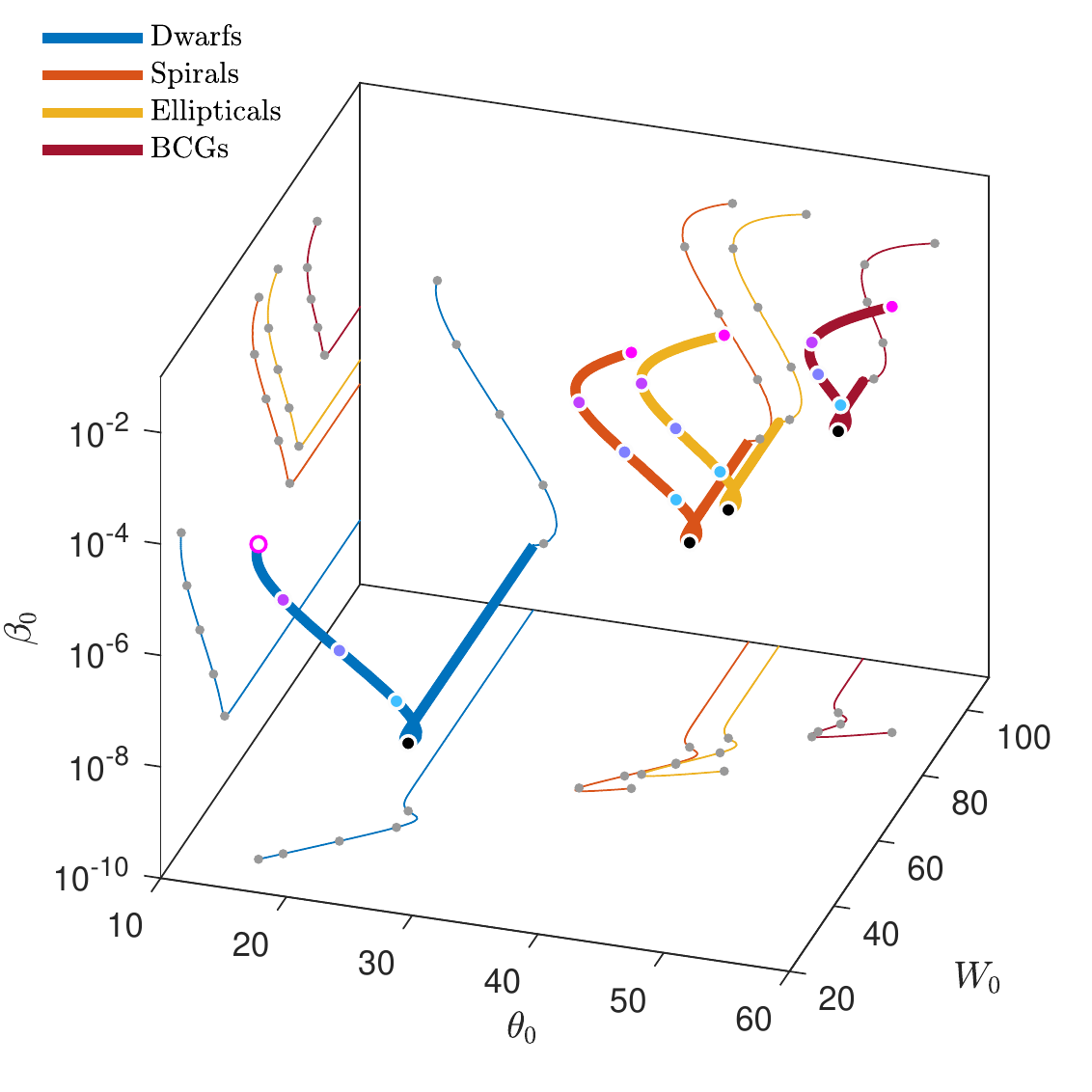}
	\caption{Full astrophysical ranges of the configuration parameters for $m c^2=\SI{48}{\kilo\eV}$. Specific observational halo constraints (\acs{Mh} and \acs{rh}) for each astrophysical case define $1$-dimensional curves in the ($\beta_0, \theta_0, W_0$)-space. Each coloured dot along each thick line has its corresponding \ac{RAR} benchmark solution (for $\acs{M-dm}(r)$, $\acs{v-dm}(r)$ and $\rho(r)$) in \cref{fig:profiles}. The limiting filled-magenta dots correspond to the critical solutions for spirals, ellipticals and BCGs while the empty-magenta dot is associated with the limiting (non-critical) solution for typical dwarfs. The latter is set by the (threshold) solution without developing a maximum in the halo rotation curve. The straight line behaviour in all cases correspond to solutions having the minimum core mass (or minimum $\rho_0$), as well as achieving the more extended density tails as can be seen from \cref{fig:profiles}. These solutions develop isothermal-like tails, ending in the standard isothermal density tail $\rho\propto r^{-2}$ for infinitely large cut-off parameter $W_0$.}%
	\label{fig:betathetaWZero}%
\end{figure*}

\subsection{Limiting behavior in the parameter sets}
\label{sec:analysis-2}
The effects of the \ac{RAR} parameter sets in the corresponding \ac{RAR} solutions explain the limiting values in the core mass \acs{Mc}, the total mass \acs{Mtot} and plateau density \acs{rhop}. The maximum and minimum \ac{DM} masses, predicted by the \ac{RAR} model, have associated maximum and minimum in the predicted \acs{rhop} values, as explicitly shown in \cref{fig:ParameterAnalysis}.

The importance of those predicted windows for each galaxy type reflects universal relations between galaxy parameters. Thus, the predicted windows of \acs{Mc} and \acs{Mtot} masses reflect the Ferrarese Universal relation \citep{2002ApJ...578...90F} while the predicted \acs{rhop} window reflects the constancy of the \textit{central} surface \ac{DM} density in galaxies \citep{2009MNRAS.397.1169D}. Both relations are discussed in detail in \cref{sec:correlations}.

The main responsible for the increase of the quantum-core mass, i.e. from $\sim\SI{E3}{\Msun}$ in dSphs to $\sim\SI{E8}{\Msun}$ in typical spirals, ellipticals and BCGs, is the temperature parameter $\beta_0$, which can vary about six orders of magnitude among the different galaxy types. Instead, the pair ($\theta_0$, $W_0$) remains around the same order-of-magnitude values and is mainly relevant to the \ac{DM} halo physics. For the latter compare \cref{fig:profiles} and \cref{fig:betathetaWZero}, together with values in \cref{tbl:profiles}.

The temperature parameter $\beta_0$ is limited from above by its critical value \acs{beta0-crit} for the case of typical spiral and elliptical galaxies as well as BCGs. That limit is set by the \textit{quantum core} condition. For higher values the \ac{RAR} solutions become gravitationally unstable and lead to the gravitational collapse of the quantum core. In the case of typical dwarf galaxies the temperature is limited by its maximum value \acs{beta0-max}. That limit, on the other hand, is set by the (threshold) solution without a maximum in the halo rotation curve, corresponding to highly cuspy halos. Thus, while $\beta_0^{\rm cr}$ sets the critical core mass \acs{Mc-crit} for typical spiral and elliptical galaxies as well as BCGs, the \acs{beta0-max} sets the \acs{Mc-max} for typical dSphs. See \cref{fig:ParameterAnalysis} and \cref{tbl:profiles} for numerical values for each galaxy type.

At the same time, a specific minimal temperature parameter \acs{beta0-min} (for all galaxy types and BCGs) is implied by the linear relation between the configuration parameters ($\beta_0$, $\theta_0$, $W_0$), as seen in \cref{fig:betathetaWZero} and the corresponding projection-planes (for $\beta_0$ not close to its maximum). For large enough $W_0$ values (and beyond) the solutions develop isothermal halo tails without affecting the inner structures through surface effects. Especially, the core remains constant, what resembles here mainly constant $\beta_0$ and $\theta_0$ values. Thus, large enough $W_0$ values set all the possible total halo masses \acs{Mtot}, although unbounded from above because $W_0$ may grow up to infinity. Correspondingly those solutions imply the minimal temperature \acs{beta0-min} which produce minimal \ac{DM} core masses (see the $W_0,\beta_0$ projection plane).

The existence of a \acs{beta0-crit} for spirals, ellipticals and BCGs (and \acs{beta0-max} for dwarfs), will necessarily define through the above monotonic relation a low enough $W_0$ value to set the minimal total halo mass for each galaxy type. In the case of spirals, ellipticals and BCGs that minimum correlates with the maximal (critical) temperature \acs{beta0-crit}. The correlation does not apply for dwarfs due to the strong boundary effects (i.e. small-sized) close to the maximal temperature \acs{beta0-max}, see \cref{fig:ParameterAnalysis}.

\section{Robustness of the \ac{RAR} model predictions}
\label{sec:robustness}
\def\ROOTPATH{figures/dwarfs-robustness}\begin{figure}[t!]%
	\centering%
	\includegraphics[width=\hsize]{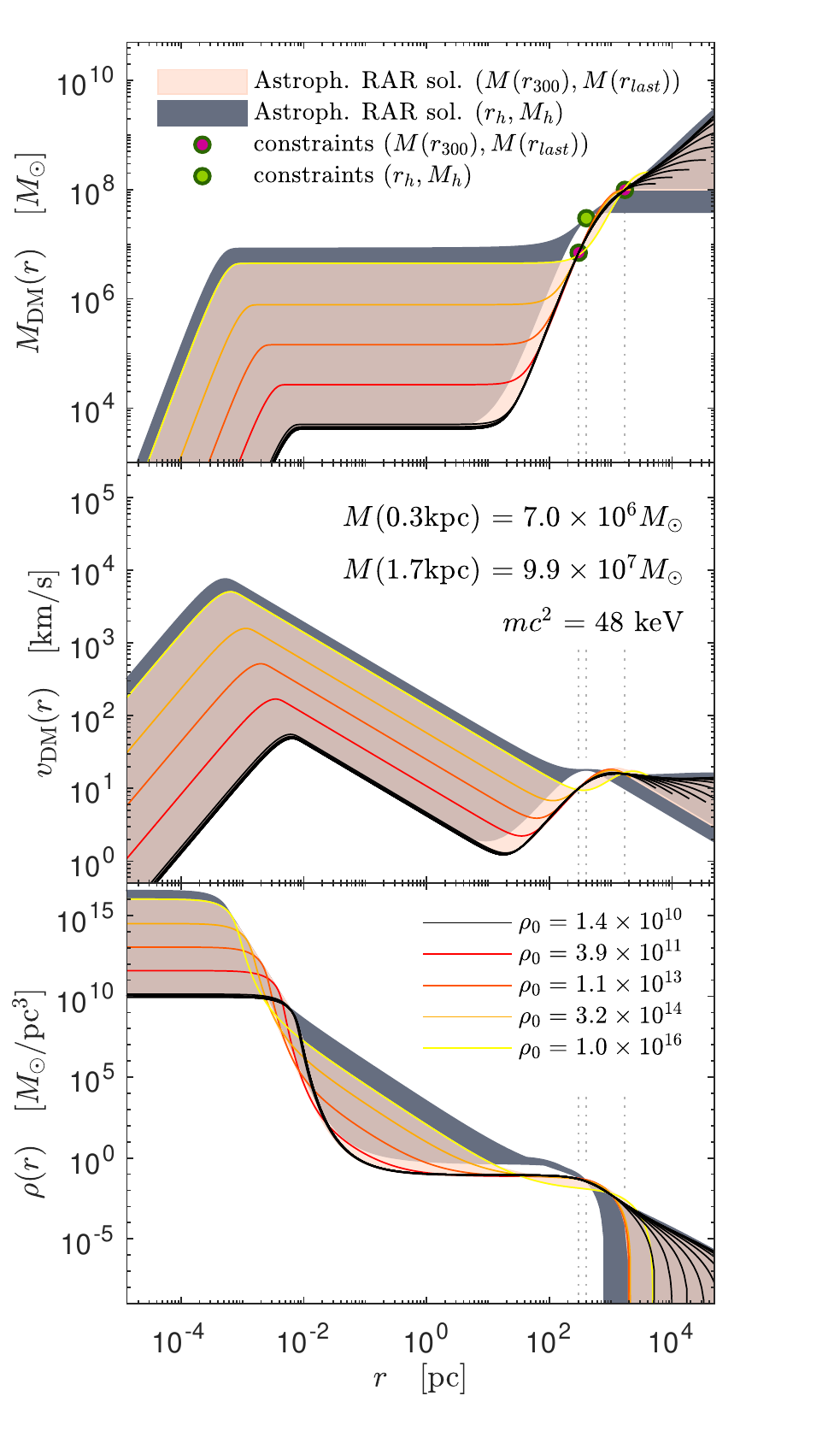}
	\caption{Comparison between the full window of the astrophysical \ac{RAR} solutions, as obtained by using the (\acs{rh}, \acs{Mh}) halo constraints (gray shaded regions, coinciding with the bluish region in \cref{fig:profiles}), with those fulfilling the alternative halo constraints ($M(\acs{r-300}),M(\acs{r-last})$), represented by the light-red shaded regions. The latter belongs to the Fornax dwarf as given in \citet{2009ApJ...704.1274W}. There is a mild shift in the position of the maximum-circular velocities (roughly a factor $2$) between both kind of families (see text for more details). The maximal velocity $\acs{v-max} \approx \SI{18}{\kilo\metre\per\second}$, on the other hand, is exactly obtained in both cases.}%
	\label{fig:profiles-dwarfs-last}%
\end{figure}
\def\ROOTPATH{figures/dwarfs-robustness}\begin{table*}
	\includegraphics[width=\hsize]{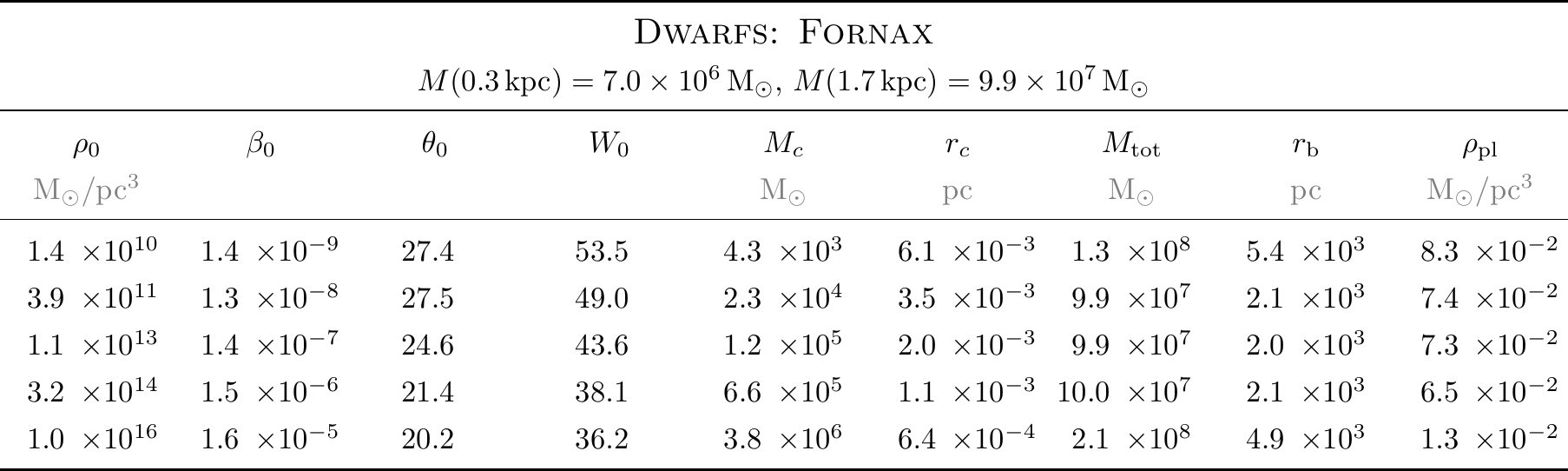}
	\caption{Free \ac{RAR} model parameters (with $mc^2=\SI{48}{\kilo\eV}$) for the $5$ benchmark sets of \ac{DM} profiles, as shown in \cref{fig:profiles-dwarfs-last}. They fulfill the observational constraints ($M(\SI{0.3}{\kilo\parsec})=\SI{7.0E6}{\Msun}$, $M(\SI{1.7}{\kilo\parsec})=\SI{9.9E7}{\Msun}$) for the Fornax dwarf as given in \citet{2009ApJ...704.1274W}. The characteristic DM masses and radii, given here , have to be compared with those given in \cref{tbl:profiles} for typical dSphs. Importantly, the predicted \acs{Mc}, \acs{Mtot} and \acs{rhop} values, by the RAR model in this case, are very similar to the ones obtained for the case of the other halo constraints (\acs{rh}, \acs{Mh}). This indicates the robustness in the RAR predictions.}
	\label{tbl:fornax}
\end{table*}
The allowed choice for observational constraints at the rotation curves maxima (\acs{rh}, \acs{Mh}), are here selected in order to have a convenient and unique prior to be used across the entire galaxy zoo. Nevertheless, more precise observational halo mass constraints can be obtained at other typical radial halo scales (though somewhat close to \acs{rh}), depending on the galaxy type. For example, in the case of dSphs, the halo mass is observationally better constrained at \acs{r-300} (i.e at $\SI{300}{\parsec}$, very close to \acs{r-half} for Milky Way satellites) as shown in \citet{2008Natur.454.1096S}. Including spiral and elliptical galaxies, other typical one-halo scale lengths (such as the Burkert halo scale-length) are appropriate as reported in \citet{2009MNRAS.397.1169D}.

With the aim to analyze the robustness of the \ac{RAR} model predictions, we further investigate which are the effects on the free \ac{RAR} model parameters when changing the halo constraints $(\acs{rh}, \acs{Mh})$ to the (observationally) better constrained couple $(M(\acs{r-300}), M(\acs{r-last}))$. Here, \acs{r-last} is the last observed data point, as reported in \citet{2009ApJ...704.1274W}, allowing for a good fit of $M(\acs{r-last})$. These constraints represent the case of typical dSphs.

The results show a mild shift between the new set of astrophysical \ac{RAR} solutions, illustrated as a light-red shaded region in \cref{fig:profiles-dwarfs-last}, with respect to the one already found in \cref{fig:profiles} (displayed as grey-shaded region in \cref{fig:profiles-dwarfs-last}). Correspondingly, we found similar sets of free \ac{RAR} free model parameters for the new benchmark solutions, as explicited in \cref{tbl:fornax}, which should be compared with those in \cref{tbl:profiles}.

It is worth to note that with the new set of constraints our results predict a narrower range for the core mass, $\acs{Mc} \approx \SIrange{4.3E3}{3.8E6}{\Msun}$, compared to the result in \cref{sec:results}. This is an increase of about $\SI{10}{\percent}$ for the lower limit and a decrease of about $\SI{46}{\percent}$ for the upper limit. The total mass shows an offset towards higher values with increase of about $\SIrange{200}{250}{\percent}$.

Similarly, the same results predict also a narrower range for the surface density $2/3\acs{rhop}\acs{rh} \approx \SIrange{19.5}{78}{\Msun/\parsec^2}$, being well within the $3\sigma$ uncertainty area as shown \cref{fig:donato}. Note that there is an decrease in the plateau density \acs{rhop} of about one order of magnitude but also an increase in the halo radius \acs{rh} of about the same order of magnitude. In sum the product $\acs{rhop}\acs{rh}$ remains robust towards a change of constraints.

The main conclusions from the alternative constraints are very similar to conclusions from the halo constraints (\acs{rh}, \acs{Mh}). Thus, we obtain similar effects on the predicted \ac{DM} magnitudes (such as \acs{Mc}, \acs{Mtot} and \acs{rhop}) for the differently chosen boundary halo conditions. This maintains intact the main predictions as provided through the halo constraints (\acs{rh}, \acs{Mh}).
\section[The halo radius vs. halo mass RAR relation]{The \acs{rh} - \acs{Mh} RAR relation}
\label{sec:rhMh-relation}

The aim here is to show that by the usage of the nearly constant DM surface density $\acs{rho0D} \acs{r0} \approx \SI{140}{\Msun\parsec^{-2}}$ (best-fit) as given in \citet{2009MNRAS.397.1169D}, an approximate (but very useful) relation between the one-halo scale length of the RAR solutions \acs{rh} and its mass \acs{Mh}, is possible (with \acs{rho0D} and \acs{r0} the \textit{central} \ac{DM} halo density at the one-halo-scale-length of the Burkert profile). Such (\acs{rh},\acs{Mh}) relation, obtained without the need to go for specific galaxy-type observables as done in \cref{sec:constraints}, will be thus applied to cover the gap between dSphs and elliptical galaxies in \cref{fig:McMDM} (represented by the green-ish area).   

We start the deduction by recalling the relation between Burkert and RAR one-halo scale lengths $\acs{r0}\approx 2/3 \acs{rh}$, as well as the identification $\acs{rho0D}\equiv\acs{rhop}$, as given in \cref{sec:correlations:donato}. This allows us to write the following new version of the Donato relation

\begin{equation}
\label{eqn:DonatoRAR}
\frac23\acs{rhop}\acs{rh} = \SI{140}{\Msun/\parsec^2}
\end{equation}

Given that typical RAR family solutions develop extended plateaus of nearly constant density all the way up to halo scales, the following approximation holds

\begin{align}
\label{eqn:rhopl-rhMh}
\acs{rhop} \approx \frac{3 \acs{Mh}}{4 \pi \acs{rh}^3}
\end{align}

\begin{figure}%
	\centering%
	\includegraphics[width=\hsize]{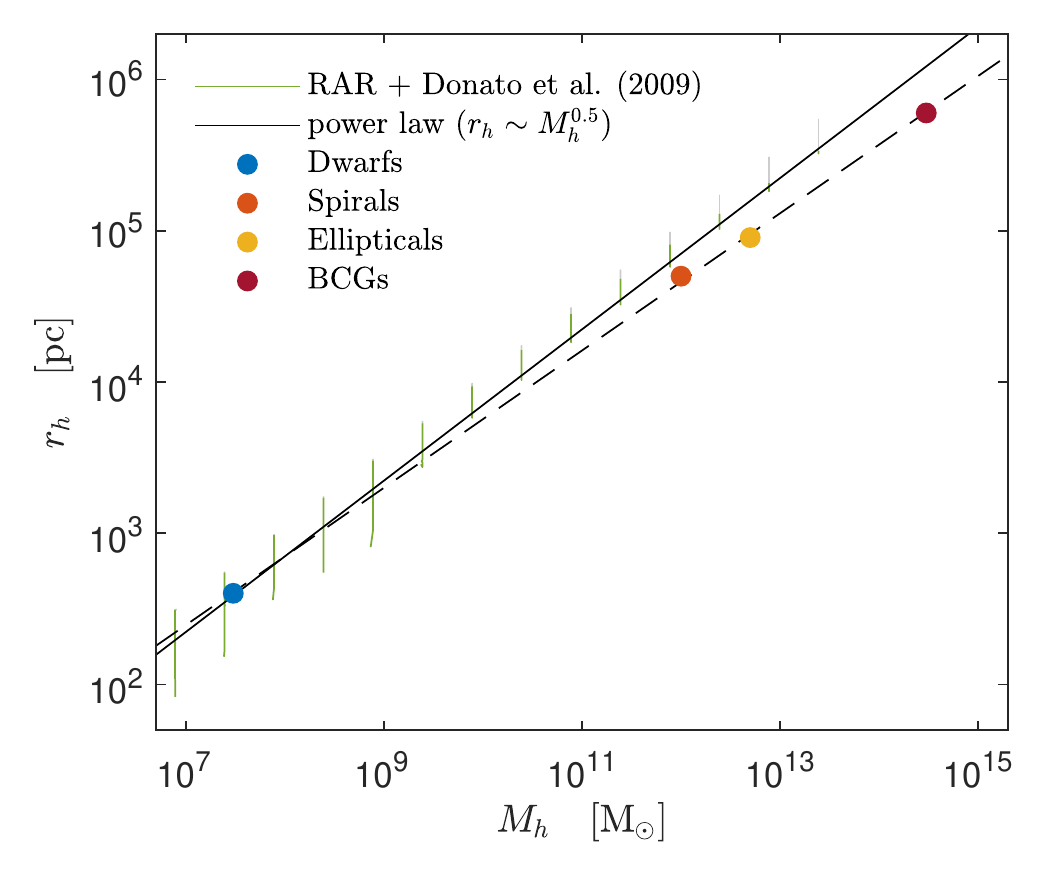}
	\caption{In dashed (power law) line: correlation between \acs{rh} and \acs{Mh} as inferred from typical galactic structures in \cref{sec:constraints}, where the coloured points show the typical values of dwarfs, spirals, ellipticals and BCGs. In continuous line: the approximate power law relation $\acs{rh} \sim \acs{Mh}^{0.5}$ as obtained by using the Donato (best-fit) constant value. The vertical dotted lines are RAR predictions for a given halo mass (corresponding with the white curves in \cref{fig:McMDM}) for the fixed mean surface density $\frac23\acs{rhop}\acs{rh} = \SI{140}{\Msun/\parsec^2}$ given by Donato in \citet{2009MNRAS.397.1169D}.}%
	\label{fig:MhRh}%
\end{figure}

The combination of the above two equations directly implies a power law relation $\acs{rh} \sim \acs{Mh}^{0.5}$, which is plotted in \cref{fig:MhRh}, and compared with the somewhat milder power relation followed by the observational constraints (\acs{rh},\acs{Mh}) obtained in \cref{sec:constraints} from dSph to BCGs (in dashed line joining the coloured dots). The reason why these two relations cannot coincide can be seeing through the RAR predictions (vertical thick lines) to the Donato relation in \cref{fig:donato}, which indeed do not exactly match the Donato best-fit for larger galaxy types. Therefore, an alternative (but equivalent) approach to calculate other families of \ac{RAR} profiles from new boundary conditions (\acs{rh},\acs{Mh}) (obtained here from the approximate power law) is possible. Thus, for given halo masses \acs{Mh} within the range $\SIrange{E7}{E12}{\Msun}$ (and for a fixed surface density $\frac23\acs{rhop}\acs{rh} = \SI{140}{\Msun/\parsec^2}$), one obtains corresponding halo radius \acs{rh}, depending on the model parameters. The results cover a regime between dwarfs and ellipticals, shown in the green-ish area in \cref{fig:McMDM}, where the white (benchmark) lines (labeled by \acs{Mtot} in the horizontal regime) show the new set of RAR families obtained for those new (\acs{rh},\acs{Mh}) values. Notice that the continuation of the predicted green-ish area is not extended up to BCGs because the constancy of the DM surface density by \citet{2009MNRAS.397.1169D} was only provided up to ellipticals.

\def\ROOTPATH{figures/MhRh}

\bibliography{biblio}

\end{document}